\documentclass[aps,pra,twocolumn,amsmath,amssymb,showpacs]{revtex4-1}
\usepackage{graphicx}
\usepackage{amsmath}
\usepackage{bm}

\newcommand{\be}{\begin{equation}}  
\newcommand{\ee}{\end{equation}}
\newcommand{\ba}{\begin{array}}
\newcommand{\ea}{\end{array}}
\newcommand{\bea}{\begin{eqnarray}}
\newcommand{\eea}{\end{eqnarray}}

\newcommand{\bra}{\langle}
\newcommand{\ket}{\rangle}

\newcommand{\nn}{\nonumber}

\begin{document}

\title{
Energetic and entropic effects of bath-induced coherences 
} 

\author{C.L. Latune$^1$, I. Sinayskiy$^{1,2}$, F. Petruccione$^{1,2,3}$}
\affiliation{$^1$Quantum Research Group, School of Chemistry and Physics, University of
KwaZulu-Natal, Durban, KwaZulu-Natal, 4001, South Africa\\
$^2$National Institute for Theoretical Physics (NITheP), KwaZulu-Natal, 4001, South Africa\\
$^3$School of Electrical Engineering, KAIST, Daejeon, 34141, Republic of Korea}

\date{\today}
\begin{abstract}
The unavoidable interaction of a quantum system with its surrounding (bath) is not always detrimental for quantum properties. For instance, under some specific conditions (that we identify as {\it indistinguishability}), a many-body system can gain internal coherences thanks to the interaction with its bath. The most famous consequence of this phenomenon is superradiance. Beyond that, the thermodynamic effects on the system of these bath-induced coherences have been mostly unexplored. We show here, for a simple and common system (a pair of two-level systems), that the energetic and entropic impacts can indeed be dramatic and diverse, including amplification of the action of the bath but also its mitigation. Our results can be tested experimentally. They suggest that bath-induced coherences can be harnessed to enhance thermodynamic tasks, opening up interesting perspectives for thermal machines, quantum battery charging, natural or artificial energy harvesting systems, and state preparation and protection. 
\end{abstract}


\maketitle
\section{Introduction}
The fundamental importance of quantum coherences in quantum thermodynamics has been shown in a growing number of problems ranging from thermal machines \cite{Scully_2003,Zhang_2007,Dillenschneider_2009,Scully_2011, Rahav_2012, Dorfman_2013, Brandner_2015, Uzdin_2015, Niedenzu_2015, Gelbwaser_2015, Leggio_2015b, Mitchison_2015, Killoran_2015, Korzekwa_2016, Uzdin_2016, Chen_2016, Su_2016,Turkpence_2016, Dag_2016, Niedenzu_2016, Mehta_2017, Dag_2018, Levy_2018, Xu_2018, Holubec_2018,Wertnik_2018}, quantum battery charging \cite{Campaioli_2017,Ferraro_2018,Campaioli_2018}, heat flow \cite{apptemppaper}, energy transport \cite{Caruso_2009}, photovoltaic energy conversion \cite{Scully_2010}, and photosynthesis \cite{Romero_2014}.   
Still, the role of coherences and the full extend of its impact is far from being fully understood. 
 Beyond its use, the production, manipulation and conservation of coherences represent serious challenges particularly due to the unavoidable influence of baths which tends to leave any system in thermal states (with no coherences left to use or extract). 
However, under some specific conditions (to be detailed), the interaction with the bath becomes beneficial for coherences. It is a well-known phenomenon, brilliantly exploited in superradiance \cite{Dicke_1954, Gross_1982} and generation of entanglement \cite{Benatti_2003, Benatti_2010, Passos_2018}. Recently, a study \cite{Cakmak_2017} focused specifically on the amount of coherences that can be generated through bath interaction.  

Here, aiming at using bath-induced coherences for thermodynamic applications, we focus on alternative aspects, mostly unexplored so far, namely the energetic and entropic impacts for the system of bath-induced coherences. 
We show that the effects can indeed be diverse and drastic, and persist for most initial states and any bath temperature. As a special case we study initial thermal states, which are particularly important for thermodynamics and experiments.
We find that for a pair of two-level systems, one of the simplest and experimentally accessible system exhibiting bath-induced coherences, phenomena of mitigation of the bath effects can happen, resulting in large reduction or increase of both the steady state energy and entropy of the system. 
Even more interesting, phenomena of amplification of the bath effects can also happen, implying as well a large reduction or increase of the system's steady state energy. 

 The underlying phenomenon responsible for such effects is identified as the indistinguishability of the subsystems from the point of view of the bath. 
Focusing on a pair of two-level systems presents several advantages like 
allowing for simple experimental realisations and providing the opportunity to better understand the consequences of bath-induced coherences before dealing with more complex systems. 

The phenomena of bath mitigation due to bath-induced coherences was already pointed out in \cite{apptemppaper}. Here, we extend the analysis to arbitrary initial states (instead of only the ground state), negative effective bath temperatures (which emerges for instance when two baths interact with the system \cite{Brunner_2012}, or in autonomous thermal machines \cite{autonomous}), and include also entropy considerations. Thus, the scope of the results is greatly increased and the phenomenon of amplification of the bath effects is uncovered. The relation between energy amplification/attenuation, entropy amplification/attenuation, and bath-induced coherences is made explicit.  
An apparent paradox emerges since coherences do not contribute to the energy so that the common sense tells us that coherences are not able to affect the energy of a system. An intuitive explanation of this paradox is provided in the light of the apparent temperature introduced in \cite{apptemppaper}.
Our results have promissing applications in quantum thermodynamics, thermal machines, and quantum battery charging, but also potential applications for natural or artificial energy harvesting devices, and state preparations or protection for computation and quantum error correction.

\section{Model}
We consider a pair of two-level systems (two-level atoms or spins $1/2$) of energy transition $\omega$ ($\hbar=1$) and interacting with a bath in a thermal state at inverse temperature $\beta_B$. Importantly, we considered that $\beta_B$ can be positive or negative since the combination of several baths or systems can result in an effective bath having some transitions (pair of levels) with negative temperatures \cite{Brunner_2012}. If such transitions are resonant with the two-level systems, the effective bath behaves formally as a bath at negative temperature. Similarly, the quantum battery of an autonomous thermal machine reaches a steady state with an (apparent) temperature which can take negative values \cite{autonomous}, as if the evolution of the quantum battery was driven by a thermal state in a negative temperature.  

Furthermore, we assume that the interaction with the bath gives raise to {\it collective} dissipation. We explain in the following what we mean by collective dissipation and what are the underlying conditions. Taking the seminal example of two-level atoms interacting with the free space electric field 
(which can be extended straightforwardly to spins), the interaction is of the form (under the dipole approximation) \cite{Gross_1982}
\be\label{gencoupling}
V= -\sum_{i=1,2} \vec D_i.\vec E(\vec r_i),
\ee
where $\vec E(\vec r_i)$ is the electric field operator at the position $\vec r_i$ of the atom $i$, and $\vec D_i = d(\sigma_i^{+} + \sigma_i^{-})\vec\epsilon_i$ is the dipole operator of the atom $i$ with $d$ the electric dipole (identical for both atoms) and $\vec\epsilon_i$ the polarisation of the atomic transition between the ground state $|0\ket_i$  to the excited state $|1\ket_i$. The operators $\sigma_i^{+}=|1\ket_i\bra 0|$ and $\sigma_i^{-}=|0\ket_i\bra 1|$ are the ladder operators of the atom $i$, realising such transition. The above interaction Hamiltonian \eqref{gencoupling} can be rewritten in the form,
\be\label{gencoupling2}
V= -\sum_{i=1,2} (\sigma_i^{+} + \sigma_i^{-}) B_i
\ee
where $B_i := d \vec \epsilon_i.\vec E(\vec r_i)$ corresponds to the bath operator interacting with the atom $i$. 
If the two atoms are far apart (typically separated by a distance much larger than the emission wavelength $\lambda_a = c/\omega$), one can show that the expectation value $\langle B_1 B_2 \rangle_{\rho_{\rm bath}} :={\rm Tr} B_1 B_2 \rho_{\rm bath}$ is equal to zero (assuming that the electromagnetic field is in a thermal state, see more detail in \cite{Gross_1982} and Appendix \ref{apppol}). Thus, one can consider that each atom interacts effectively with their own independent bath represented by the operators $B_1$ and $B_2$.
 Alternatively, if the atoms have orthogonal polarisations $\vec \epsilon_1.\vec \epsilon_2 =0$ (while confined in a volume much smaller than the emission wavelength $\lambda_a$), one can show (Appendix \ref{apppol}) that the bath operators $B_i$  still satisfy $\langle B_1 B_2 \rangle_{\rho_{\rm bath}} =0$ implying again that each atom interacts effectively with its own independent bath. 
 
 Under such conditions and assuming a weak bath coupling so that the Markov and Born approximations \cite{Cohen_Book, Petruccione_Book} are valid, the dissipative dynamics of the system is given by a master equation of the form \cite{Gross_1982}
\bea\label{medis}
\dot{\rho}_S^I &=& -i\Omega_L [\sigma_1^{+}\sigma_1^{-}+\sigma_2^{+}\sigma_2^{-},\rho_S^I] \nn\\
&&+ g [n(\omega) +1]\sum_{i=1}^2 (2\sigma_i^-\rho_S^I \sigma_i^+ -\sigma_i^+\sigma_i^-\rho_S^I - \rho_S^I\sigma_i^+\sigma_i^-)\nn\\
&&+ g n(\omega) \sum_{i=1}^2(2\sigma_i^+\rho_S^I \sigma_i^- -\sigma_i^-\sigma_i^+\rho_S^I - \rho_S^I\sigma_i^-\sigma_i^+),\nn\\
\eea
where $n(\omega)$ is the bath mean excitation number at the frequency $\omega$,
$\Omega_L$ is the Lamb shift, and $g= \frac{d^2\omega^3}{6\pi c^3\hbar \epsilon_0}$ is the effective coupling strength with $c$ the vacuum light velocity and $\epsilon_0$ is the vacuum permeability.
The above master equation corresponds to two atoms interacting with their own independent bath of same characteristics (temperature and coupling strength).  \\

The opposite situation is when the atoms are at the same position (or at least confined in a volume much smaller than the emission wavelength $\lambda_a = c/\omega$) and with parallel polarisation, $\vec \epsilon_1=\vec \epsilon_2$, so that the two bath operator $B_1$ and $B_2$ are equal: the two atoms are effectively {\it indistinguishable} to the bath. 
 In such a situation the dissipation is given by the following dynamics \cite{Gross_1982,Cakmak_2017} (still assuming a weak coupling with the bath),
 \bea\label{meind}
\dot{\rho}_S^I &=& -i\Omega_L [\sigma_1^{+}\sigma_1^{-} + \sigma_2^{+}\sigma_2^{-},\rho_S^I]-i\Omega_{I}[\sigma_1^{+}\sigma_{2}^{-}+\sigma_1^{-}\sigma_{2}^{+},\rho_S^I] \nn\\
&&+ g [n(\omega) +1] (2S^-\rho_S^I S^+ -S^+S^-\rho_S^I - \rho_S^IS^+S^-)\nn\\
&&+ g n(\omega) (2S^+\rho_S^I S^- -S^-S^+\rho_S^I - \rho_S^IS^-S^+),\nn\\
\eea
where $S^{\pm} = \sigma_1^{\pm} +\sigma_2^{\pm}$ are the {\it collective} ladder operators, and $\Omega_I$ characterises the interaction strength between the two atoms (which appears due to the spatial confinement of the pair, see \cite{Gross_1982} and Appendix \ref{tlatoms}). Note that the work in \cite{Cakmak_2017} studies the generation of coherences induced by the bath in a pair of two-level atoms considering also the above dynamics \eqref{medis}. However, they focus mainly on the situation where the pair is initialised in the ground state (and on the temporary impact of the atom interaction $\Omega_I$ for some specific initial states). Moreover, all the considerations on the energetic and entropic impact on the steady state of the pair, which constitute the main contributions of our paper, are absent of \cite{Cakmak_2017}.  \\

In the following, we refer to the dynamics described by \eqref{medis} as {\it independent dissipation}. As shown above, it results from the different position or polarisation of the two atoms. In other words, the two atoms bear different individual characteristics which makes them {\it distinguishable} from the point of view of the bath. This concept can be extended to any system. We therefore call {\it distinguishable} any pair (or larger ensemble) of subsystems bearing different individual characteristics resulting in an independent interaction of each subsystem with its own effective bath (as described by Eq. \eqref{medis}).
By contrast, we refer to the dynamics \eqref{meind} as {\it collective dissipation}, which was shown to be a consequence of the same individual characteristic of the two atoms, making them {\it indistinguishable} to the bath. Therefore, we call {\it indistinguishable} any pair (or larger ensemble) of subsystems bearing the same characteristics so that each subsystem interacts with exactly the same bath (as described by Eq. \eqref{meind}). Note that indistinguishability can also be enforced through bath engineering (as for instance by inserting an optical cavity \cite{Woods_2014}).  
 The above example of a pair of two-level atoms interacting with the free space electromagnetic field is merely an illustration of these notions of distinguishability and indistinguishability. 
 In the remainder of the paper we focus on their consequences for the steady states and the associated thermodynamic properties.

\section{Steady states} 
The steady state of the independent dynamics Eq. \eqref{medis} is well-known since it corresponds to the independent dissipation of each subsystems. The steady state is therefore the product of the steady state of each subsystems, namely, thermal states at the bath temperature \cite{Petruccione_Book}, 
\bea
\rho^{\rm th}(\beta_B) &=& Z^{-1}(\beta_B) e^{-H_0\beta_B} \nn\\
&=&Z^{-1}(\beta_B) \Big[e^{-2\omega\beta_B}|1\ket|1\ket\bra 1|\bra1|\\
+&e^{-\omega\beta_B}&(|1\ket|0\ket\bra 1|\bra0|+|0\ket|1\ket\bra0|\bra1|) + |0\ket|0\ket\bra0|\bra0|\Big]\nn
\eea
(the tensor product order is taken to be the same for ``bras'' and ``kets''), where $Z(\beta_B):=1+2e^{-\omega\beta_B} + e^{-2\omega\beta_B}$, and 
\be\label{freeH}
H_0:= \omega (\sigma_1^{+}\sigma_1^{-} + \sigma_2^{+}\sigma_2^{-}),
\ee
 is the free Hamiltonian of the pair of two-level systems.\\

 The steady state of the collective dissipation, denoted by $\rho^{\rm ss}(\beta_B,r)$, can be easily found by projecting Eq. \eqref{meind} onto the basis of symmetric and anti-symmetric states $|\psi_{\pm}\ket :=(|0\ket|1\ket\pm|1\ket|0\ket)/\sqrt{2}$, $|\psi_{0}\ket := |0\ket|0\ket$, and $|\psi_1\ket:=|1\ket|1\ket$. We obtain (see Appendix \ref{dynamics}),
\bea\label{genss}
 && \rho^{\rm ss}(\beta_B,r) = (1-r)|\psi_{-}\ket\bra\psi_{-}| + rZ_{+}^{-1}(\beta_B) \\
&&\hspace{0.7cm}\times\Big(e^{-2\omega\beta_B}|\psi_1\ket\bra \psi_1|+e^{-\omega\beta_B}|\psi_+\ket\bra\psi_+| + |\psi_0\ket\bra\psi_0|\Big),\nn
\eea
where $r:= \bra \psi_0|\rho(0)|\psi_0\ket + \bra \psi_1|\rho(0)|\psi_1\ket + \bra \psi_{+}|\rho(0)|\psi_{+}\ket$ and $Z_{+}(\beta_B):= 1+e^{-\omega\beta_B} +e^{-2\omega\beta_B}$. 
The steady state of the collective dynamics contrasts largely with the thermal state $\rho^{\rm th}(\beta_B)$, which brings several observations. First, $\rho^{\rm ss}(\beta_B,r)$ is not unique and depends on the initial state through $r$, which is a striking difference from the steady state $\rho^{\rm th}(\beta_B)$. Secondly, the steady state $\rho^{\rm ss}(\beta_B,r)$ can contain non-vanishing coherences as opposed to $\rho^{\rm th}(\beta_B)$. This is the object of the following Section. \\

\section{Steady state coherences}\label{sectionsscoh}
 In this Section we show explicitly that the steady state of the collective dissipation, $\rho^{\rm ss}(\beta_B,r)$, can contain (global) coherences in the form of coherent superposition of the states $|0\ket|1\ket$ and $|1\ket|0\ket$ (corresponding to non-diagonal terms in the basis $\{|0\ket|0\ket,|1\ket|0\ket,|0\ket|1\ket,|1\ket|1\ket\}$). Such coherences are induced (or maintained, if initially present) by the bath. Roughly speaking, since the bath does not distinguish between the two-level systems, each time an excitation is absorbed from the bath or emitted to the bath, both two-level systems gain or lose simultaneously an excitation (if in the ground or excited state), generating correlations between them. 
 More precisely, during these processes of absorption and emission the transitions $|1\ket|1\ket \leftrightarrow |\psi_{+}\ket$ and $|0\ket|0\ket \leftrightarrow|\psi_{+}\ket$ take place, which involves coherent superpositions of $|0\ket|1\ket$ and $|1\ket|0\ket$ ($|\psi_{+}\ket= (|1\ket|0\ket +|0\ket|1\ket)/\sqrt{2}$), generating (or maintaining) coherences in the system. In Appendix \ref{appindistin} we detail more this idea around the role of indistinguishability and draw a parallel with entanglement generation in quantum optics. 
 
 One should note that the coherences between $|1\ket|0\ket$ and $|0\ket|1\ket$ corresponds also to correlations between the two-level systems. This can be seen be observing that the reduced steady state of each two-level system is always diagonal, so that the tensor product of the local states is different from $\rho^{\rm ss}(\beta_B,r)$, indicating the presence of correlations. Therefore, the coherences between $|1\ket|0\ket$ and $|0\ket|1\ket$ can be seen alternatively as correlations between the two two-level systems.
 
 The expression of the steady state coherences is obtained directly from the expression of the steady state \eqref{genss},
  \bea
  \bra 0|\bra1|\rho^{\rm ss}(\beta_B,r)|1\ket|0\ket &=&\bra 1|\bra0|\rho^{\rm ss}(\beta_B,r)|0\ket|1\ket \nn\\
  &=& \frac{1}{2}\left(\frac{r}{z(\beta_B)} -1\right),
  \eea
  with $z(\beta_B):= \frac{Z_+(\beta_B)}{Z(\beta_B)}$.
It is convenient for the remainder of the paper to define $c:= \bra 0|\bra1|\rho^{\rm ss}(\beta_B,r)|1\ket|0\ket + \bra 1|\bra0|\rho^{\rm ss}(\beta_B,r)|0\ket|1\ket=\left(r/z(\beta_B)-1\right)$ the sum of the steady state coherences, which can take any value within the interval $[-1;\frac{1}{3}]$. One should note that a proper measure of coherence as defined in \cite{Baumgratz_2014} is for instance the $l_1$ norm of coherence (the sum of the absolute value of each coherence), which would give here ${\cal C}_{l_1}[\rho^{\rm ss}(\beta_B,r)]= | \bra 0|\bra1|\rho^{\rm ss}(\beta_B,r)|1\ket|0\ket| + | \bra 1|\bra0|\rho^{\rm ss}(\beta_B,r)|0\ket|1\ket|= |c|$, taking value in $[0;1]$. 
Note that one recovers the result of \cite{Cakmak_2017} when the pair is initialised in the ground state by taking $r=1$, giving an amount of steady state coherence equal to ${\cal C}_{l_1}[\rho^{\rm ss}(\beta_B,1)]=1/z(\beta_B)-1$.
Nevertheless, we will see in the following that the sign of the coherences is essential, therefore the important quantity in this problem is $c$. 
   Crucially, the coherences (and therefore $c$) are strictly positive as soon as $r > z(\beta_B)$, strictly negative when $r < z(\beta_B)$, and null when $r = z(\beta_B)$. Anticipating the remainder of the paper, we stress that this observation is fundamental since, as shown in \cite{apptemppaper}, the sign of $c$ determines whether the apparent temperature \cite{apptemppaper} is increased or decreased and consequently the steady state energy (see more detail in the following).  
   
Up to now our considerations are valid for any initial state. It is however particularly interesting to look at the restricted class of initial states made of thermal states, particularly important for thermodynamics and the most common and accessible experimentally. Moreover, thermal states contain no coherence so that any coherence in the steady state is induced by the bath. Considering an initial state $\rho^{\rm th}(\beta_0)$ at the inverse temperature $\beta_0$ (allowed to be negative), the steady state is given by the general expression \eqref{genss} with $r$ equal to
\bea
r&=& \bra \psi_0|\rho^{\rm th}(\beta_0)|\psi_0\ket + \bra \psi_1|\rho^{\rm th}(\beta_0)|\psi_1\ket \nn\\
&&\hspace{2.4cm}+ \bra \psi_{+}|\rho^{\rm th}(\beta_0)|\psi_{+}\ket\nn\\
&=& z(\beta_0).
\eea
  $z(\beta_0)$ is a strictly monotonic increasing function of $\beta_0$ on $]0;+\infty]$ (from $3/4$ to 1) and decreasing on $]-\infty;0[$ (from 1 to $3/4$). In the remainder of the paper we denote by $\rho^{\rm ss}(\beta_B,\beta_0):=\rho^{\rm ss}[\beta_B,r=z(\beta_0)]$ the steady state reached by the pair when initialised in $\rho^{\rm th}(\beta_0)$. 
  
  Coming back to the considerations on the steady state coherences, we reach an enlightening conclusion. The bath-induced coherences 
   take the simple form $c=z(\beta_0)/z(\beta_B) -1$, which are strictly positive (negative) if and only if $|\beta_0|>|\beta_B|$ ($|\beta_0|<|\beta_B|$). Moreover, the only situation with no bath-induced coherences is when $|\beta_0| = |\beta_B|$. It coincides with the fact that $\rho^{\rm th}(\beta_B)$ is naturally a steady state of the collective dissipation, meaning that for $\beta_0=\beta_B$ we have $\rho^{\rm ss}(\beta_B,\beta_B) = \rho^{\rm th}(\beta_B)$ which does not contain coherences.    
The crucial role of the steady state coherences in determining the major thermodynamic properties (energy and entropy)  of the pair of two-level systems is shown in the following.\\


\begin{figure}
\centering
(a)~~\includegraphics[width=7cm, height=5cm]{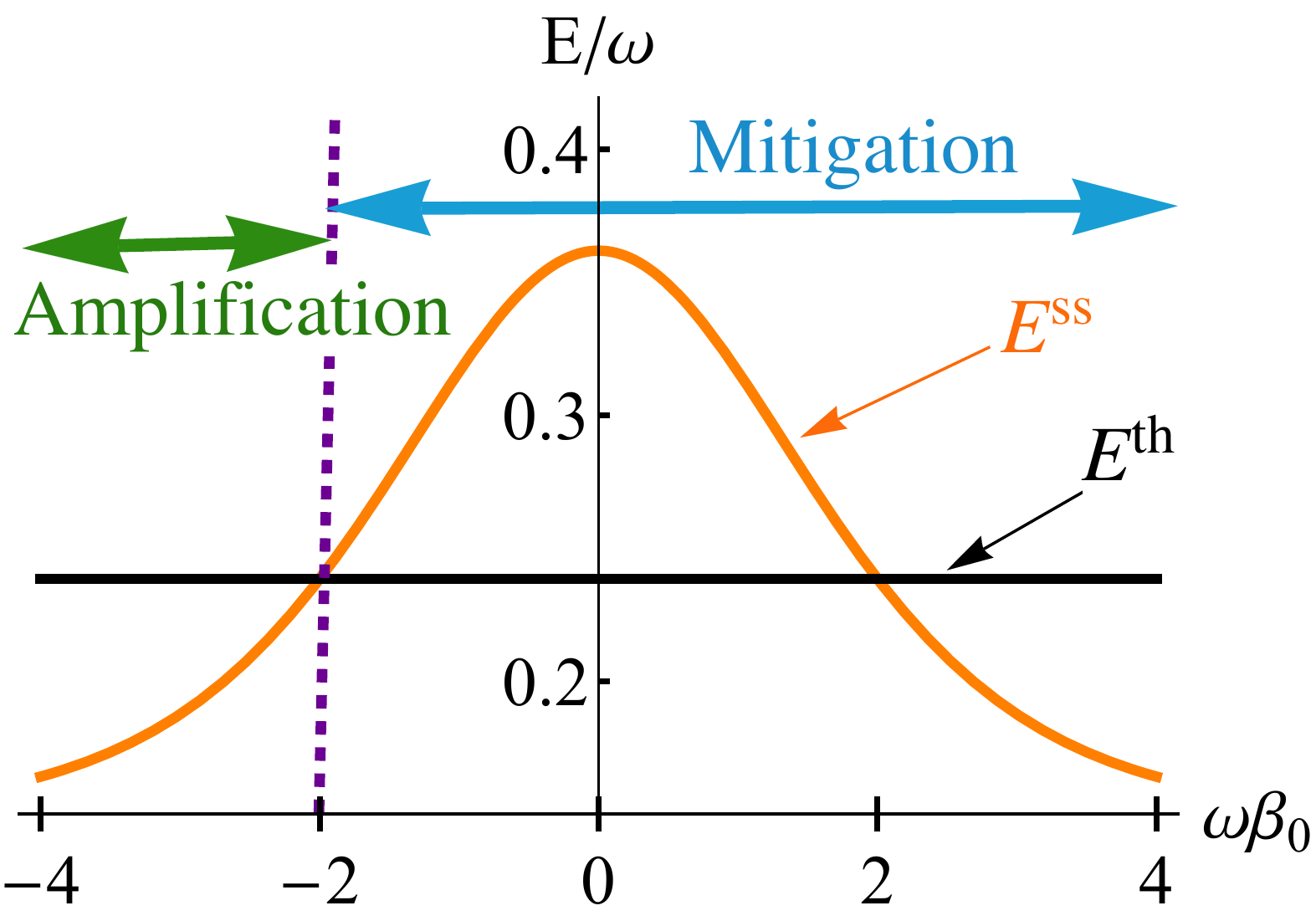}
(b)~~\includegraphics[width=7cm, height=5cm]{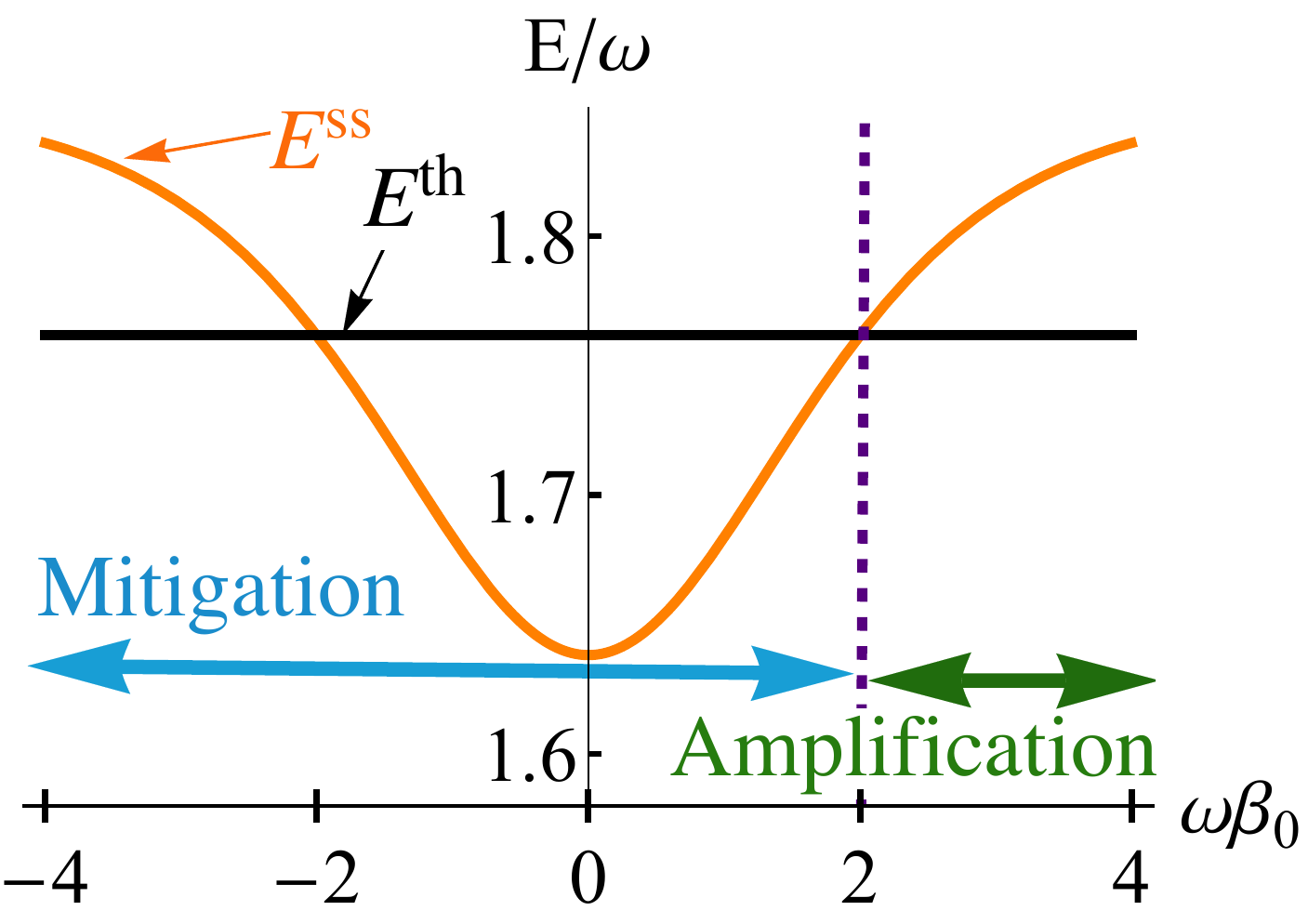}
\caption{Graph of the steady state energy $E^{\rm ss}(\beta_B,\beta_0)$ as a function of the initial inverse temperature $\beta_0$, for (a) $\omega\beta_B =2$, and (b) $\omega\beta_B=-2$. In each graph the value of the thermal energy $E^{\rm th}(\beta_B)$ is indicated as reference by the Black line.   }
\label{ssenergy}
\end{figure}

  \section{Steady state energy}\label{sectionenergy}
   The steady state energy of the collective dissipation is given by 
  \bea\label{ssen1}
  E^{\rm ss}(\beta_B,r) &: =& {\rm Tr} H_0 \rho^{\rm ss}(\beta_B,r) \nn\\
  &=& \omega \frac{r}{Z_{+}(\beta_B)} (2e^{-2\omega\beta_B} + e^{-\omega \beta_B}) +\omega(1-r),\nn\\
  \eea 
  to be compared with the thermal energy
  \be
  E^{\rm th}(\beta_B) := {\rm Tr} H_0 \rho^{\rm th}(\beta_B) = 2\omega (e^{\omega\beta_B}+1)^{-1},
  \ee
   reached by the independent dissipation. One can already see from \eqref{ssen1} that on top of a strong dependence on $r$ and consequently on the initial state, the steady state energy can be smaller or larger than the thermal energy $E^{\rm th}(\beta_B)$ (as detailed in the following).  
    Moreover, $E^{\rm ss}(\beta_B,r)$ can be simply expressed in term of the thermal energy and bath-induced coherences,
     \bea\label{energycoh}
   E^{\rm ss}(\beta_B,r) &=& E^{\rm th}(\beta_B) - \omega \frac{1-e^{-\omega\beta_B}}{1+e^{-\omega\beta_B}}c.
   \eea
     This is a rather surprising result as the coherences do not carry energy.
    How do the steady state coherences end up contributing to the steady state energy? We answer to this question in the following. First, one can notice that for positive bath temperature, $E^{\rm ss}(\beta_B,r)$ is increased with respect to the thermal energy precisely when the bath-induced coherences $c$ are negative, and conversely, $E^{\rm ss}(\beta_B,r)$ is reduced when $c$ is positive. For negative bath temperature, the above conclusions are inverted. Thus, the steady state energy $E^{\rm ss}(\beta_B,r)$ can be amplified or attenuated in a controlled way (determined by the sign of $c$ which is itself related to the initial conditions). This may have useful applications, in particular in thermodynamics and quantum thermal machines. 
    
\subsection{Mitigation and amplification of the bath effects}
In a perspective of thermodynamic applications, we consider an initial thermal state at inverse temperature $\beta_0$ as in the previous Section. Similarly with the steady state, we denote by $E^{\rm ss}(\beta_B,\beta_0):=E^{\rm ss}[\beta_B,r=z(\beta_0)]$ the steady state energy when the pair is initialised in the state $\rho^{\rm th}(\beta_0)$. We obtain the following insight. The steady state energy $E^{\rm ss}(\beta_B,\beta_0)$ is strictly larger (smaller) than $E^{\rm th}(\beta_B)$ if and only if $|\beta_0|<\beta_B$ ($|\beta_0|>\beta_B$) for positive bath temperatures, as shown in Fig. \ref{ssenergy} (a). Conversely, the steady state energy is strictly larger (smaller) than the thermal energy if and only if $|\beta_0|>\beta_B$ ($|\beta_0|<\beta_B$) for negative bath temperatures, see Fig. \ref{ssenergy} (b). Interestingly, for $\beta_0/\beta_B>-1$ (which corresponds to the regimes of parameters denoted by ``Mitigation'' in Fig. \ref{ssenergy}), the collective interaction mitigates the bath dissipation since the pair of two-level systems starts and ends with an energy larger (or smaller) than the thermal energy. 
 This might be useful for preparing or maintaining a pair in a low (or high) energy state. 
 
By contrast, when $\beta_0/\beta_B<-1$ (which corresponds to the regimes of parameters denoted by ``Amplification'' in Fig. \ref{ssenergy}), there is an amplification of the bath effects since the steady state energy goes beyond the thermal energy, meaning that the system starts with an energy larger (smaller) than $E^{\rm th}(\beta_B)$ and ends up with a steady state energy $E^{\rm ss}(\beta_B,\beta_0)$ smaller (larger) than $E^{\rm th}(\beta_B)$. Such effects can be applied to amplify a cooling process. Let's consider a pair of two-level systems undergoing a cooling process either by direct thermal contact with a cold bath at inverse temperature $\beta_B$, either with a thermal machine yielding a steady state inverse temperature $\beta_B$ \cite{autonomous} (which can also be identified as the virtual temperature of the baths acting on the thermal machine \cite{Brunner_2012}). 
Then, if the pair of two-level systems is initially in an excited state such that $\beta_0<-\beta_B<0$), the pair of systems reaches a steady state energy $E^{\rm ss}(\beta_B,\beta_0)$ strictly lower than the thermal energy $E^{\rm th}(\beta_B)$: the cooling process is amplified. 

Similarly, considering the reverse process of loading energy into the pair of two-level systems, one can consider either a direct charging by thermal contact with a bath at inverse temperature $\beta_B$, either a charging through a thermal machine yielding a steady state inverse temperature $\beta_B$. Then, when the effective inverse temperature $\beta_B$ is negative \cite{Brunner_2012, autonomous}, if the pair is initialised in a low temperature state satisfying $\beta_0>-\beta_B>0$ it reaches a steady state energy $E^{\rm ss}(\beta_B,\beta_0)$ strictly larger than the thermal energy $E^{\rm th}(\beta_B)$: the energy charging is increased. This is of great interest for quantum battery charging. 
 Remarkably, such extra performances (mitigation, super cooling and super energy charging) rely only on indistinguishability (collective interaction with the bath).

\begin{figure}
\centering
(a)\includegraphics[width=6cm, height=4cm]{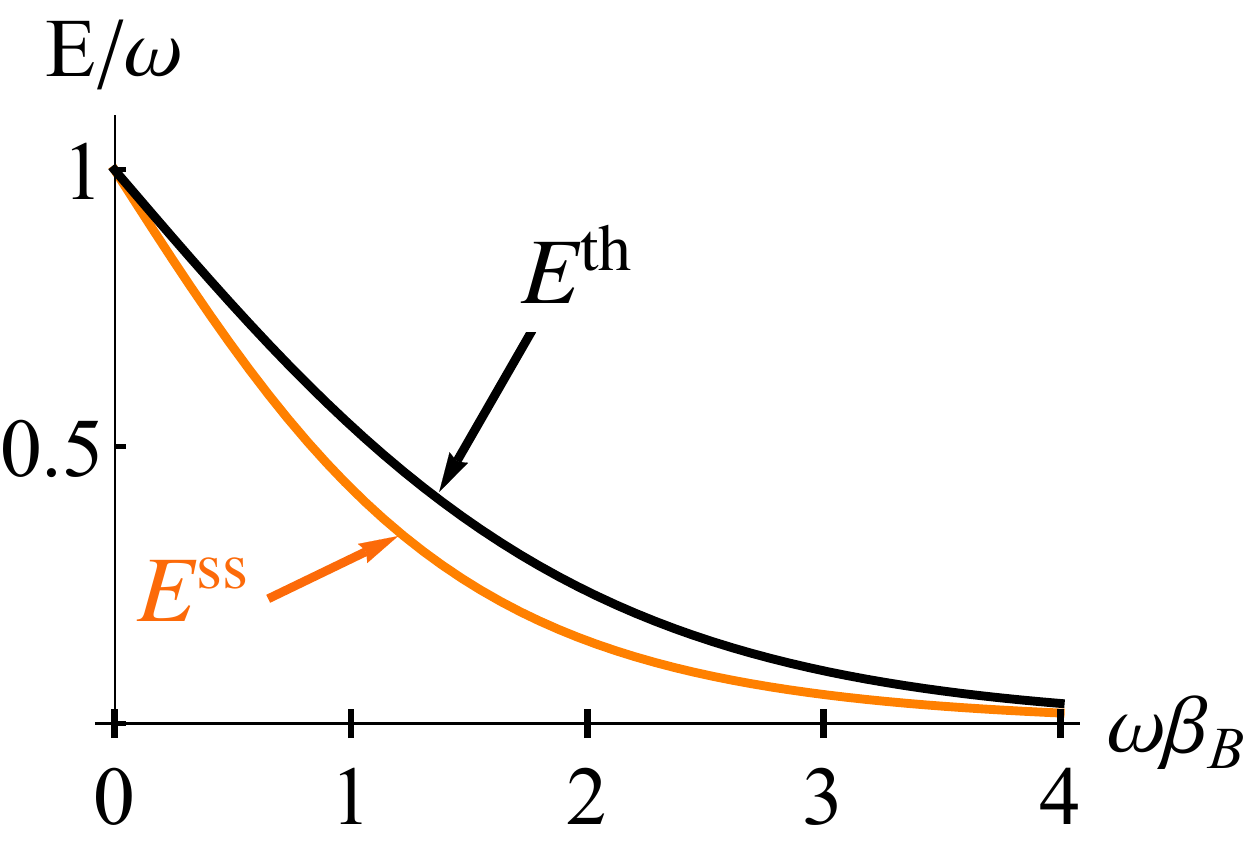}
(b)\includegraphics[width=6cm, height=4cm]{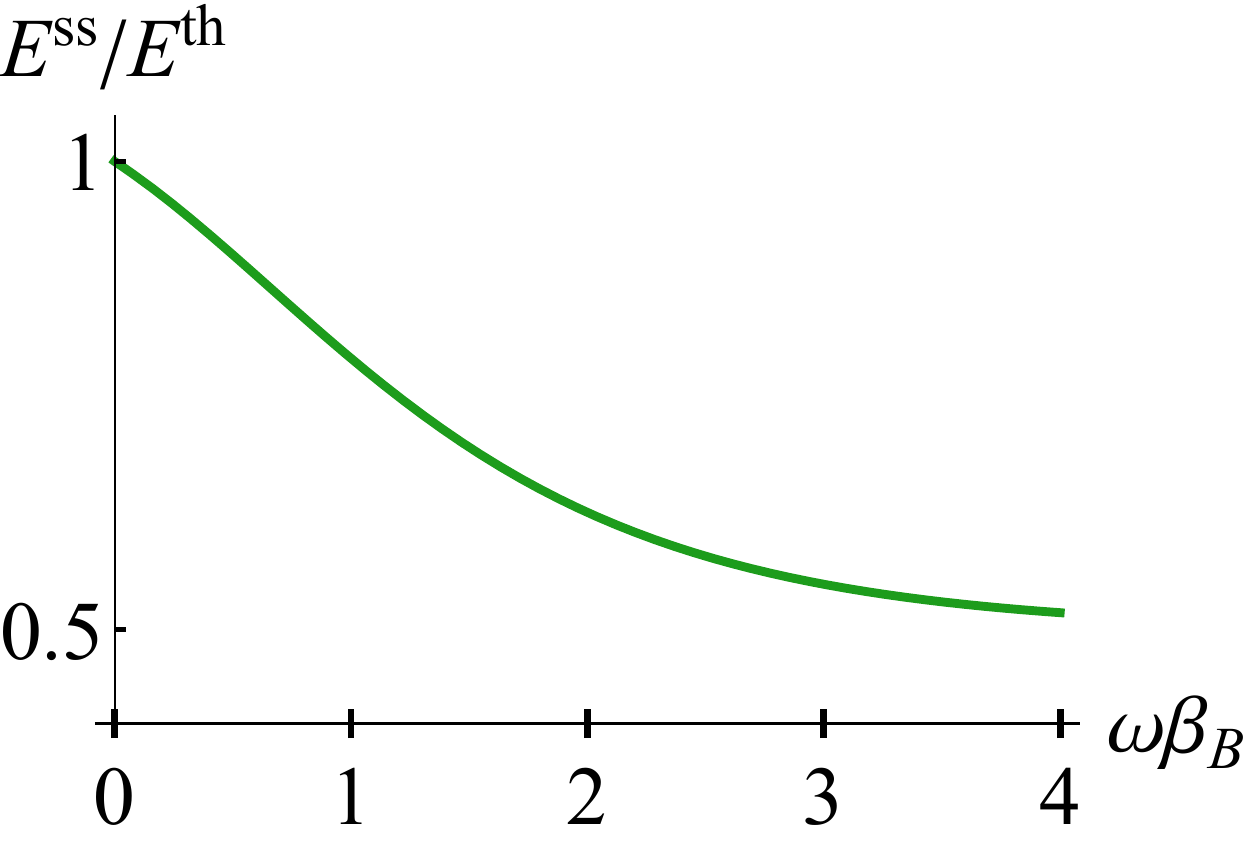}
\caption{(a) Graph of the steady state energy $E^{\rm ss}(\beta_B,\beta_0)$ (Orange curve) and thermal energy $E^{\rm th}(\beta_B)$ (Balck curve) as a function of the bath temperature $\omega \beta_B \in [0;4]$  for $\omega|\beta_0| \gg1$. (b) Corresponding ratio $E^{\rm ss}(\beta_B,\beta_0)/E^{\rm th}(\beta_B)$ as a function of $\omega \beta_B$.}
\label{enposlargeb}
\end{figure}

\begin{figure}
\centering
(a)\includegraphics[width=6cm, height=4cm]{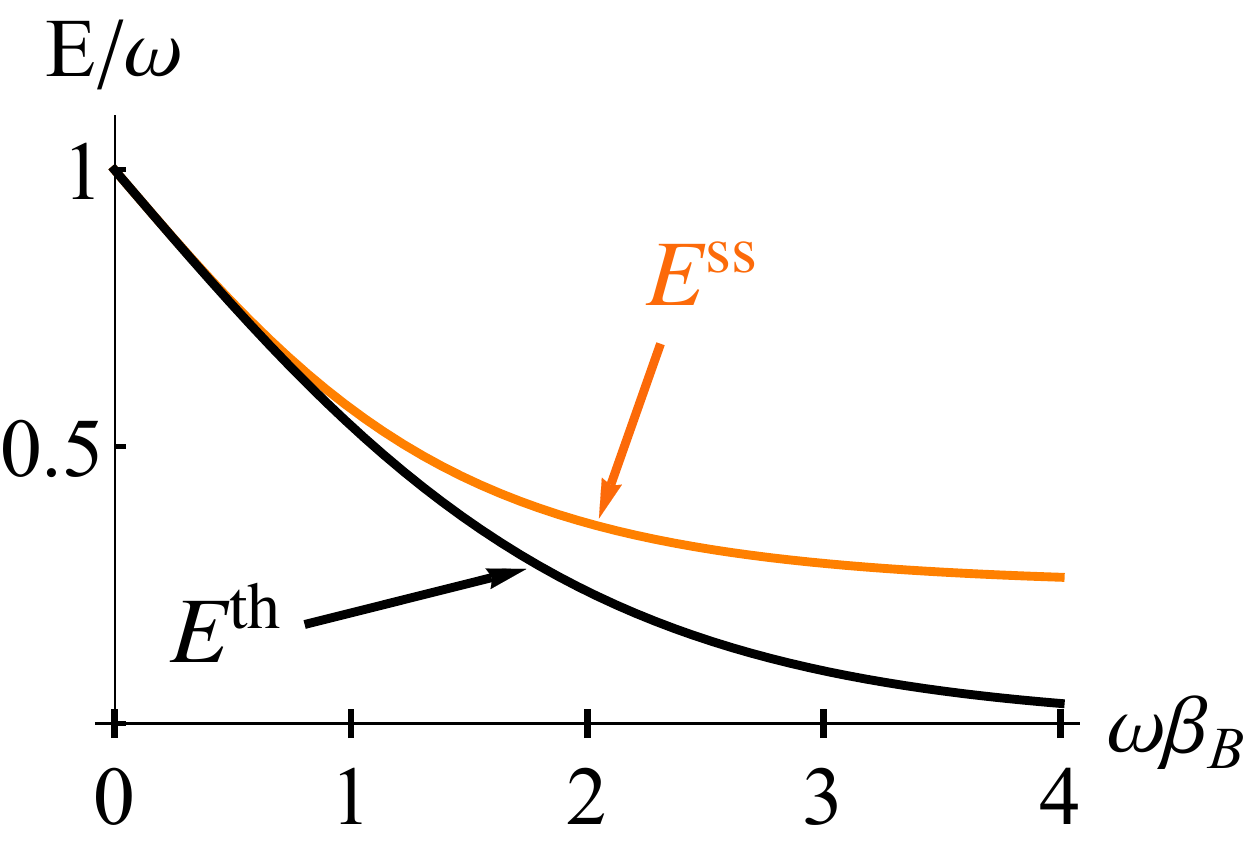}
(b)\includegraphics[width=6cm, height=4cm]{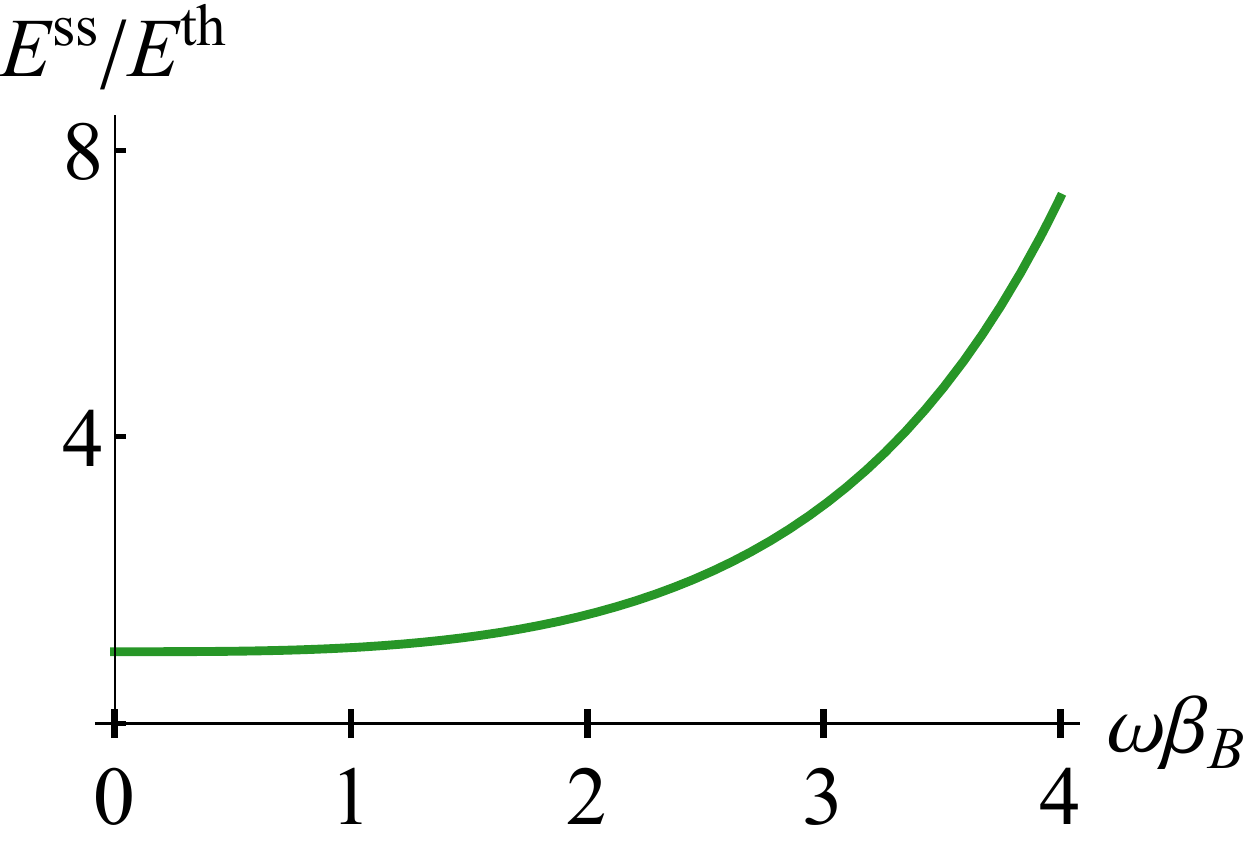}
\caption{(a) Graph of the steady state energy $E^{\rm ss}(\beta_B,\beta_0)$ (Orange curve) and thermal energy $E^{\rm th}(\beta_B)$ (Balck curve) as a function of the bath temperature $\omega\beta_B \in [0;4]$  for $\omega|\beta_0| \ll1$. (b) Corresponding ratio $E^{\rm ss}(\beta_B,\beta_0)/E^{\rm th}(\beta_B)$ as a function of $\omega\beta_B$.}
\label{enpossmallb}
\end{figure}

\begin{figure}
\centering
(a)\includegraphics[width=6cm, height=4cm]{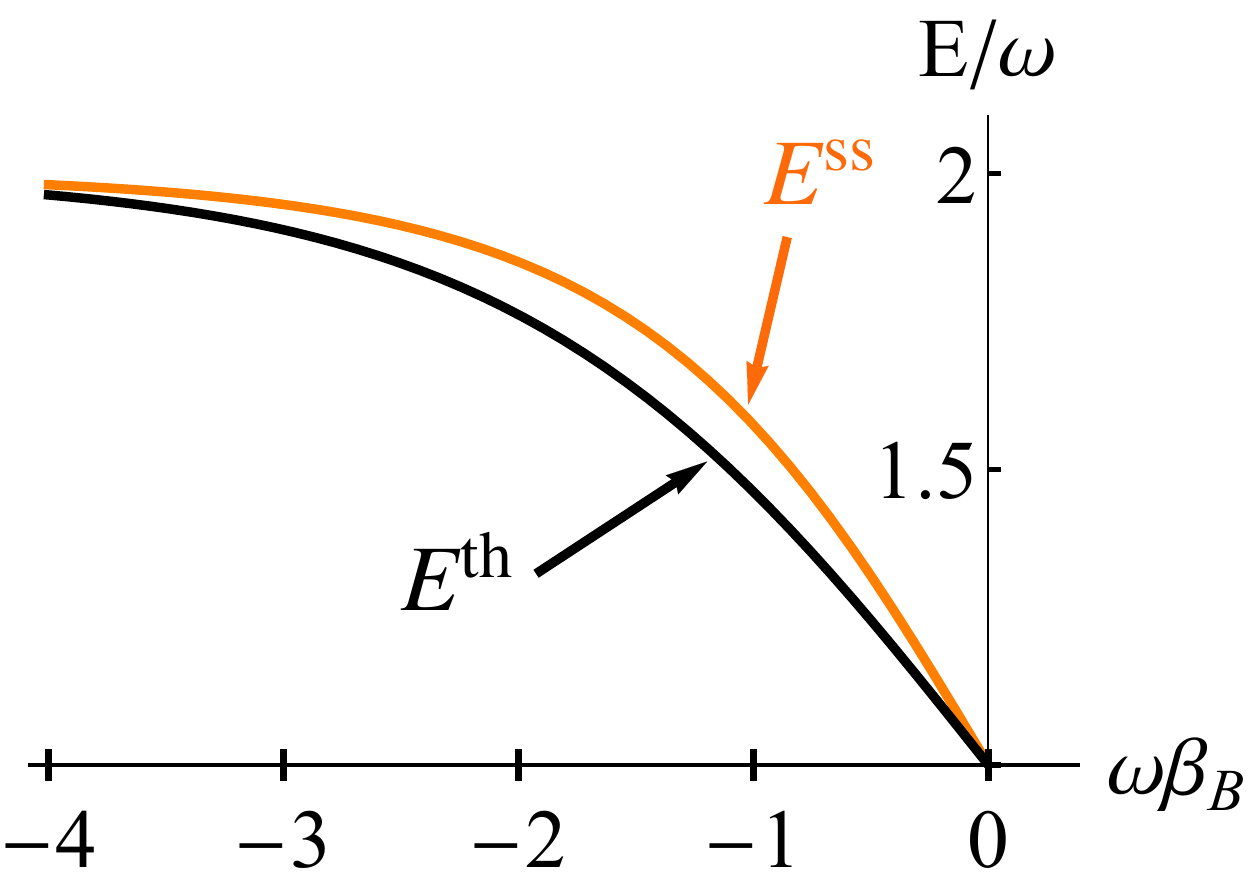}
(b)\includegraphics[width=6cm, height=4cm]{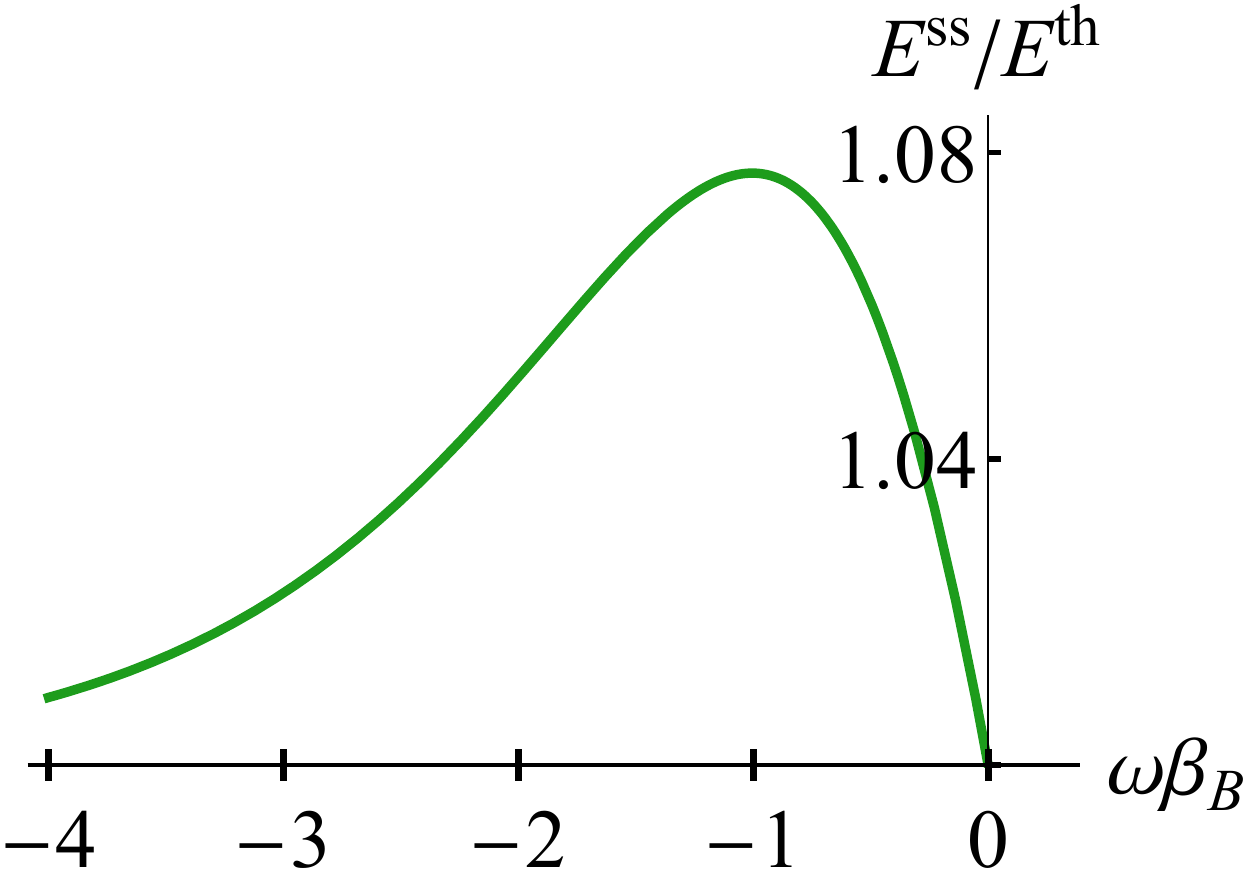}
\caption{(a) Graph of the steady state energy $E^{\rm ss}(\beta_B,\beta_0)$ (Orange curve) and thermal energy $E^{\rm th}(\beta_B)$ (Balck curve) as a function of the bath temperature $\omega \beta_B \in [-4;0]$  for $\omega|\beta_0| \gg1$. (b) Corresponding ratio $E^{\rm ss}(\beta_B,\beta_0)/E^{\rm th}(\beta_B)$ as a function of $\omega \beta_B$.}
\label{enneglargeb}
\end{figure}

\begin{figure}
\centering
(a)\includegraphics[width=6cm, height=4cm]{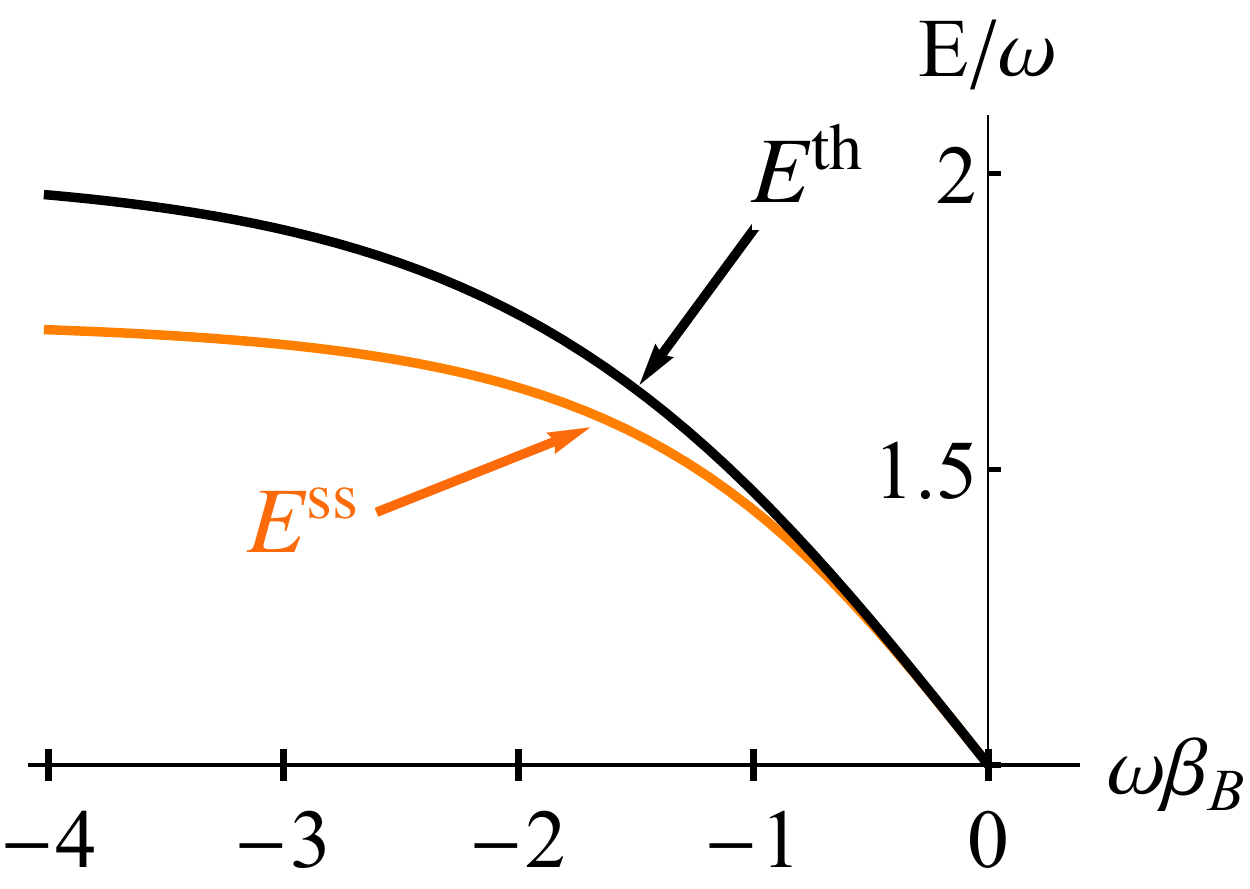}
(b)\includegraphics[width=6cm, height=4cm]{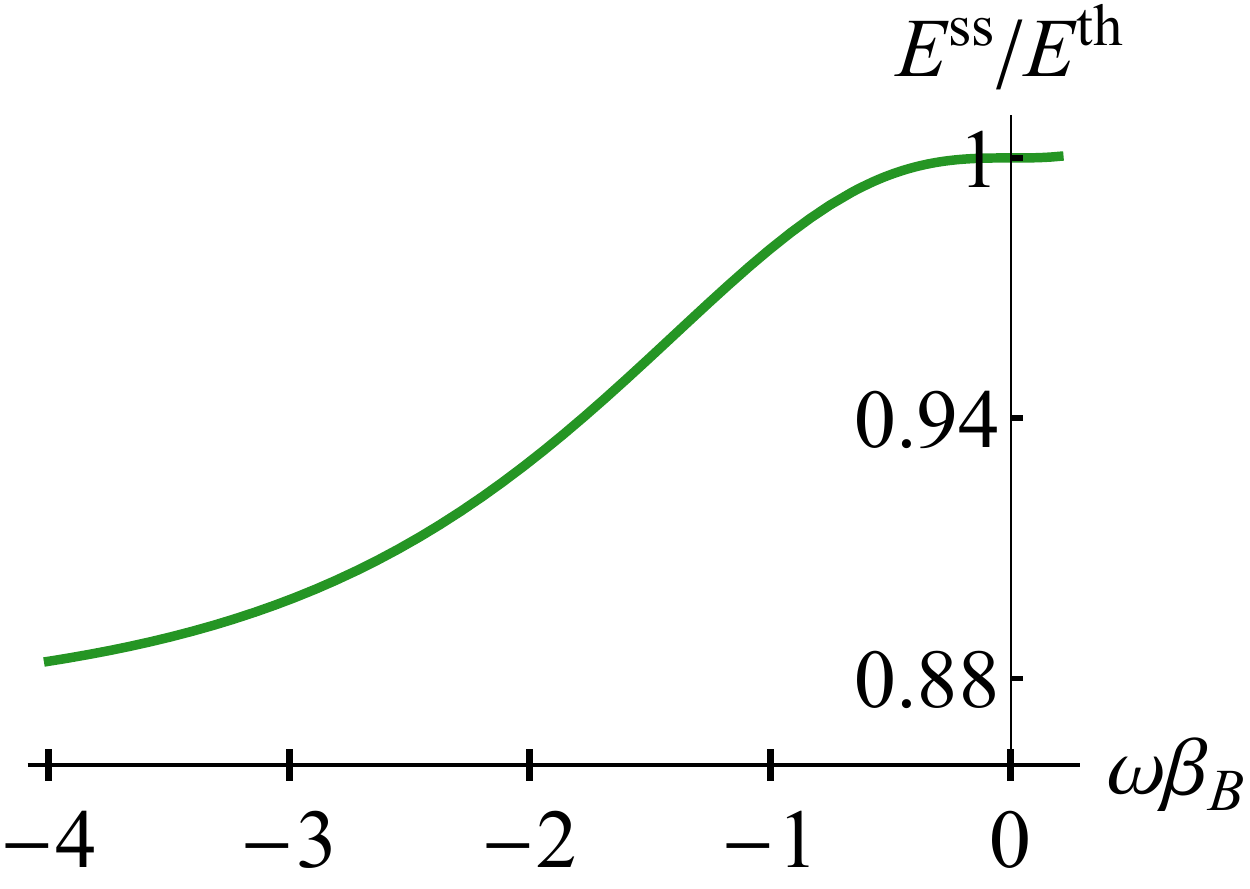}
\caption{(a) Graph of the steady state energy $E^{\rm ss}(\beta_B,\beta_0)$ (Orange curve) and thermal energy $E^{\rm th}(\beta_B)$ (Balck curve) as a function of the bath temperature $\omega \beta_B \in [-4;0]$  for $\omega|\beta_0| \ll1$. (b) Corresponding ratio $E^{\rm ss}(\beta_B,\beta_0)/E^{\rm th}(\beta_B)$ as a function of $\omega \beta_B$.}
\label{ennegsmallb}
\end{figure}

How important can be these extra performances? 
When $\beta_B>0$, the minimal value of the steady state energy $E^{\rm ss}(\beta_B,\beta_0)$ is attained for initial states such that $\omega |\beta_0| \gg 1$ and is equal to 
   \be\label{enbeta0gg1}
   E^{\rm ss}(\beta_B,\beta_0) \rightarrow_{\omega|\beta_0|\gg1} E^{\rm th}(\beta_B) - \omega \frac{1-e^{-\omega\beta_B}}{1+e^{-\omega\beta_B}}\frac{e^{-\omega\beta_B}}{Z_{+}(\beta_B)},
   \ee
which tends to be $50\%$ smaller than $E^{\rm th}(\beta_B)$ when $\omega\beta_B\gg1$. Graphs of $E^{\rm ss}(\beta_B,\beta_0)$, $E^{\rm th}(\beta_B)$, and $E^{\rm ss}(\beta_B,\beta_0)/E^{\rm th}(\beta_B)$ as functions of $\beta_B$ are represented in Fig. \ref{enposlargeb} for $\omega|\beta_0|\gg1$. Consequently, 
extra cooling of up to $50\%$ (steady state energy up to 50$\%$ smaller) can be achieved due to bath-induced coherences (indistinguishability). 
 
Conversely, the maximal steady state energy, still in presence of a positive bath temperature, is attained for $\omega|\beta_0|\ll1$, and is equal to
  \bea\label{enbeta0ll1}
   &&E^{\rm ss}(\beta_B,\beta_0) \rightarrow_{\omega|\beta_0|\ll1} E^{\rm th}(\beta_B) \\
    &&\hspace{3.5cm}-\omega \frac{1-e^{-\omega\beta_B}}{1+e^{-\omega\beta_B}}\left(\frac{3}{4}\frac{Z(\beta_B)}{Z_{+}(\beta_B)}-1\right),\nn
   \eea
   which tends to the value $\omega/4$ for $\omega \beta_B\gg1$ whereas $E^{\rm th}(\beta_B)$ tends to 0 for such bath temperatures: strong mitigation effect. Graphs of $E^{\rm ss}(\beta_B,\beta_0)$, $E^{\rm th}(\beta_B)$, and $E^{\rm ss}(\beta_B,\beta_0)/E^{\rm th}(\beta_B)$ as functions of $\beta_B$ are represented in Fig. \ref{enpossmallb} for $\omega|\beta_0|\ll1$.

%
%
%
%
%

   The above observations are inverted when the bath temperature is negative. Namely, the minimal steady state energy is attained for $\omega|\beta_0|\ll1$, and the maximal steady state energy for $\omega|\beta_0|\gg1$, with the same respective expression \eqref{enbeta0ll1} and \eqref{enbeta0gg1}. The corresponding graphs of $E^{\rm ss}(\beta_B,\beta_0)$, $E^{\rm th}(\beta_B)$, and $E^{\rm ss}(\beta_B,\beta_0)/E^{\rm th}(\beta_B)$ as functions of $\beta_B$ are represented in Fig. \ref{enneglargeb} and Fig. \ref{ennegsmallb} for $\omega|\beta_0|\gg1$ and $\omega|\beta_0| \ll1$, respectively.
    Then, in the amplification regime, corresponding to an initial thermal state such that $\beta_0 > |\beta_B|$, the bath-induced coherences can yield an extra load of energy of up to $8\%$ (Fig. \ref{enneglargeb}). \\ 

\subsection{Local temperature}
One more aspect strengthening the above considerations is that locally, each two-level systems ends up in a {\it thermal state} at an inverse temperature $\beta_{\rm Loc}$ {\it different} from the bath temperature. This can be seen directly by tracing out one of the two-level systems in $\rho^{\rm ss}(\beta_B,\beta_0)$.  We do not provide here a detailed description of the properties of $\beta_{\rm Loc}$ since it is intimately related to the ones of $E^{\rm ss}(\beta_B,\beta_0)$. We only mention briefly that the mitigation and amplification effects appear also strongly in the steady state local temperature (reaching high levels up to $33\%$ of relative increase or reduction with respect to $\beta_B$). More details can be found in Appendix \ref{localtemp}.

\subsection{The fundamental role of bath-induced coherences}
We now come back to the question raised above: how coherences, which do not carry energy, can end up contributing to the steady state energy?
  Firstly, as a preliminary observation, the heat flow between the pair of the two-level systems in a state $\rho$ 
   and the bath is characterised by the apparent temperature of the pair defined by  \cite{apptemppaper}
\be\label{defapptemp}
{\cal T} :=\omega \left(\log\frac{{\rm Tr}S^{-}S^{+}\rho}{{\rm Tr}S^{+}S^{-}\rho}\right)^{-1}.
\ee
Note that if the pair is in the dark state $|\psi_-\ket\bra \psi_-|$ it does not interact with the bath and therefore there is no heat flow and no apparent temperature can be defined. The apparent temperature ${\cal T}$ determines the direction of the heat flow in the same way as usual temperature does \cite{apptemppaper}: if ${\cal T} > 1/\beta_B$, the heat goes from the pair to the bath. Conversely, if ${\cal T} < 1/\beta_B$, the heat flows from the bath to the pair.
We can conclude that a necessary condition for a steady state is to have an apparent temperature equal to the reservoir temperature $1/\beta_B$ (otherwise the heat flow is not null). Indeed, one can verify (Appendix \ref{smapptempss}) that all states of the form \eqref{genss} have an apparent temperature equal to the bath temperature $1/\beta_B$. 

    Secondly, coherences (correlations) are built up (or maintained) due to indistinguishability, as explained above in Section \ref{sectionsscoh}. Such coherences affect dramatically the apparent temperature \cite{apptemppaper}. More precisely, for positive bath temperature, positive steady state coherences (when $r>z(\beta_B)$) increase the apparent temperature of the pair of two-level systems. Thus, the steady state of the pair cannot have the same excited state populations as the thermal state $\rho^{\rm th}(\beta_B)$ otherwise its apparent temperature would be strictly larger than the bath temperature $1/\beta_B$. Consequently, the steady state must have lower excited state populations than the thermal state $\rho^{\rm th}(\beta_B)$, which implies lower energy. 
       
    Conversely, negative coherences (corresponding to $r<z(\beta_B)$) reduce the apparent temperature so that the steady state of the pair must have higher excited state populations than the thermal state $\rho^{\rm th}(\beta_B)$ in order to reach an apparent temperature equal to the bath temperature $1/\beta_B$. The above results are inverted for negative bath temperature, namely positive (negative) coherences decrease (increase) the apparent temperature. Again, if the steady state populations are affected, the steady state energy too.
    Therefore, for these reasons, the amount of steady state coherences affect indirectly the steady state energy, as observed in Eq. \eqref{energycoh}. 
     
\subsection{Dark states}
Alternatively, the above effects can be understood in terms of dark states.         
 Considering that the pair of two-level systems is initially in a thermal state $\rho^{\rm th}(\beta_0)$ at inverse temperature $\beta_0$, it can be decomposed as a balanced combination of bright and dark states, $|\psi_+\ket$ and $|\psi_-\ket$, respectively (plus the contributions of $|\psi_0\ket$ and $|\psi_1\ket$). The dark state component is $e^{-\omega\beta_0}/Z(\beta_0)$, which is a monotonic decreasing function of $\beta_0 \in [0;\infty]$. Then, when $S$ thermalises with the bath at $\beta_B$, only the components $|\psi_0\ket$, $|\psi_+\ket$, and $|\psi_1\ket$ ``thermalise''. If $\beta_B$ is smaller than $\beta_0$, the pair ends up with a deficit of dark state component in comparison to a thermal state at inverse temperature $\beta_B$. Therefore, remembering that the dark states bears an energy $\omega$, the pair sees its energy reduced in relation to the thermal energy $E^{\rm th}(\beta_B)$. Conversely, if $\beta_B$ is larger than $\beta_0$, the pair ends up with a surplus of dark state component in comparison to a thermal state at inverse temperature $\beta_B$. Consequently, the energy of the pair $E^{\rm ss}(\beta_B,\beta_0)$ is found to be higher than the thermal energy $E^{\rm th}(\beta_B)$. This is mitigation of the bath effects. 
 
 The above rough reasoning misses some subtleties related to the normalisation factors, but this does not change the central fact that a deficit (surplus) of dark state reduces (increases) the energy of the pair. Similar considerations can be extended to negative temperatures, remembering that $e^{-\omega\beta_0}/Z(\beta_0)$ is monotonic increasing on $[-\infty;0]$, and that a deficit (surplus) of dark state component increases (reduces) the energy of the pair in comparison to the thermal energy $E^{\rm th}(\beta_B)$ for $\beta_B<0$.

\section{Steady state entropy}
 We look now at the von Neumann entropy of the steady state. 
Computing the von Neumann entropy of the steady state \eqref{genss} we have,
\bea
&&S^{\rm ss}(\beta_B,r) := S[\rho^{\rm ss}(\beta_B,r)] = -{\rm Tr} \rho^{\rm ss} (\beta_B)\log \rho^{\rm ss} (\beta_B) \nn\\
&&\hspace{1.4cm}= -r \log r -(1-r)\log(1-r)\nn\\
&& \hspace{0.4cm}+ r \log Z_{+}(\beta_B) + \omega\beta_BrZ_{+}^{-1}(\beta_B)(e^{-\omega \beta_B}+2e^{-2\omega \beta_B}) \nn\\
&&\hspace{1.4cm}= -r \log r -(1-r)\log(1-r)\nn\\
&&\hspace{0.4cm} + r \log Z_{+}(\beta_B) + \beta_B E^{\rm ss}(\beta_B) - \omega\beta_B(1-r),
\eea
to be compared with the thermal state entropy,
\bea
S^{\rm th}(\beta_B) &:=& S[\rho^{\rm th}(\beta_B)] \nn\\
&=&\log Z(\beta_B) + \beta_B E^{\rm th}(\beta_B).
\eea
The steady state entropy $S^{\rm ss}(\beta_B,r)$ can be re-written in terms of $S^{\rm th}(\beta_B) $ and $c$,
\bea\label{entropyc}
S^{\rm ss}(\beta_B,r) &=& S^{\rm th}(\beta_B) -(1-r)\log[1-e^{\omega\beta_B}Z_{+}(\beta_B)c]\nn\\
&& -r\log(1+c)-\omega\beta_B\frac{1-e^{-\omega\beta_B}}{1+e^{-\omega\beta_B}} c.
\eea
The above relation shows 
the impact of the steady state coherences on the steady state entropy. 
As a preliminary observation, one recovers that when the steady state coherences are null ($c=0$), $S^{\rm ss}(\beta_B,r)=S^{\rm th}(\beta_B)$. However, for $c\ne0$, $S^{\rm ss}(\beta_B,r)$ can be amplified or attenuated depending on the steady state coherences and initial conditions, which again can have interesting applications for state preparation, quantum thermal machines and in thermodynamics. More precisely, one can show (see Appendix \ref{appentropy}) that when $c>0$, $S^{\rm ss}(\beta_B,r)< S^{\rm th}(\beta_B)$. By contrast, for $c<0$, there is a critical value $c^{*}<0$ such that $S^{\rm ss}(\beta_B,r)> S^{\rm th}(\beta_B)$ for $c \in ]c^{*};0[$, and $S^{\rm ss}(\beta_B,r)< S^{\rm th}(\beta_B)$ for $c \in [-1;c^{*}[$ (see Appendix \ref{appentropy}).

\begin{figure}
\centering
\includegraphics[width=6cm, height=4cm]{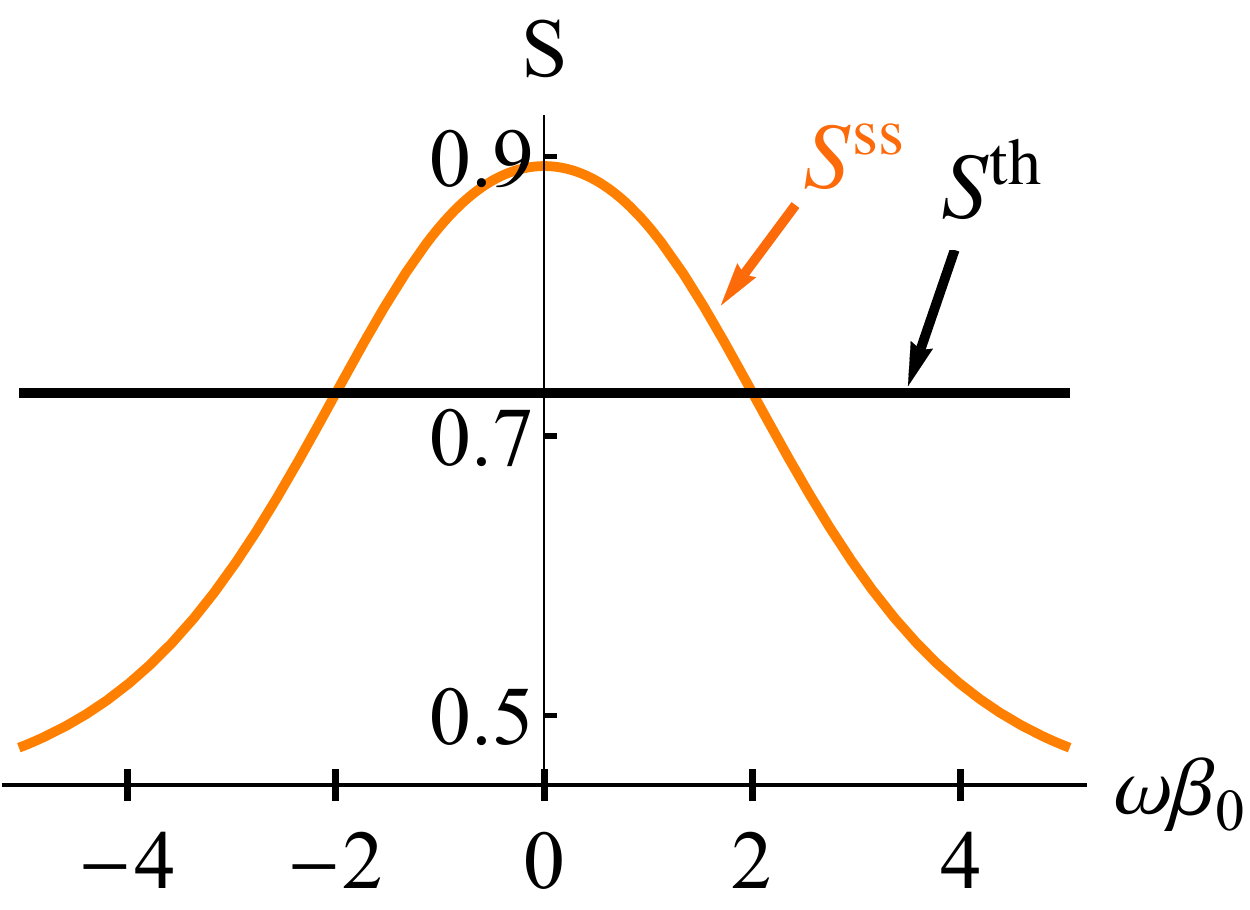}
\caption{Graphs of $S^{\rm ss}(\beta_B,\beta_0)$ (Orange curve) as a function of $\beta_0$ for $\omega\beta_B=2$ (or equivalently $\omega \beta_B=-2$). The corresponding value of $S^{\rm th}(\beta_B)$ is indicated by the Black line for comparison.}
\label{entropbb}
\end{figure}

 As previously, for thermodynamic and experimental purposes we consider the special situation where the pair of two-level systems is initially in a thermal state at the inverse temperature $\beta_0$, and we denote by $S^{\rm ss}(\beta_B,\beta_0):=S^{\rm ss}[\beta_B,r=z(\beta_0)]$ the corresponding steady state entropy. It comes the following elegantly simple result. The initial condition $|\beta_0|>|\beta_B|$ (equivalent to $S^{\rm th}(\beta_0)< S^{\rm th}(\beta_B)$) implies  $S^{\rm ss}(\beta_B,\beta_0)< S^{\rm th}(\beta_B)$, and conversely, $|\beta_0|<|\beta_B|$ (equivalent to $S^{\rm th}(\beta_0)> S^{\rm th}(\beta_B)$) implies $S^{\rm ss}(\beta_B,\beta_0)> S^{\rm th}(\beta_B)$, see Fig. \ref{entropbb}. Then, differently from the energy,  the collective dissipation has always a mitigating effect for the entropy.
 
\begin{figure}
\centering
(a)\includegraphics[width=4cm, height=3cm]{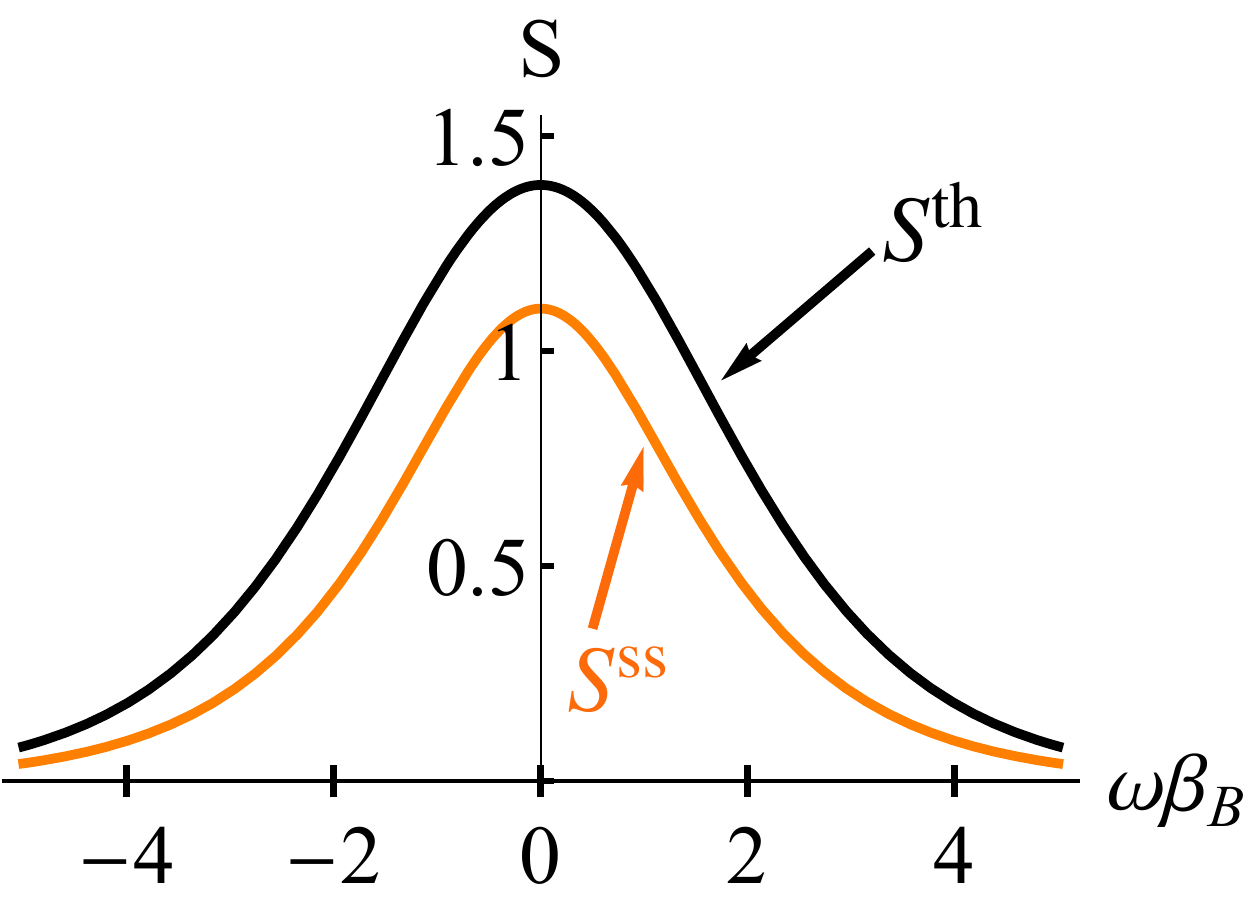}\includegraphics[width=4cm, height=3cm]{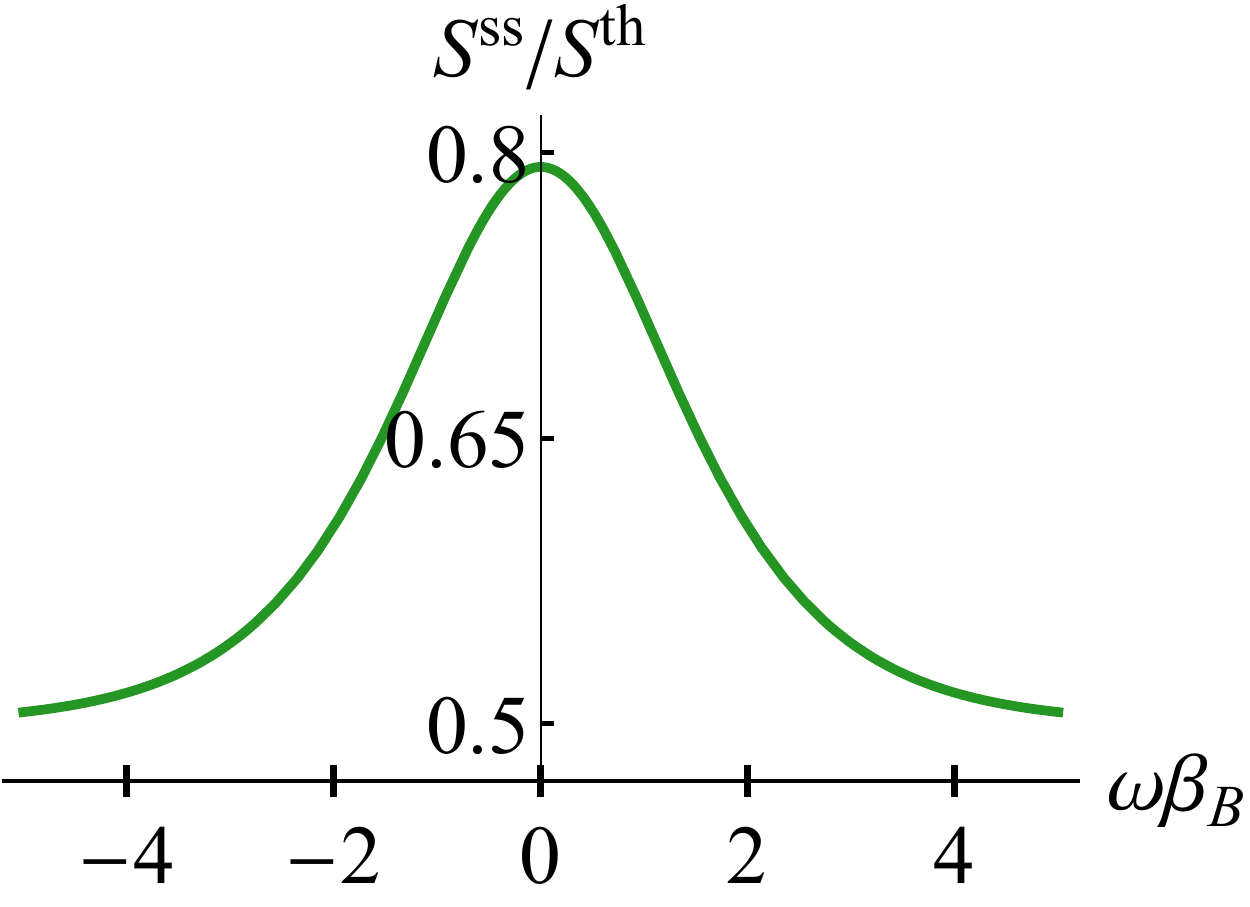}
(b)\includegraphics[width=4cm, height=3cm]{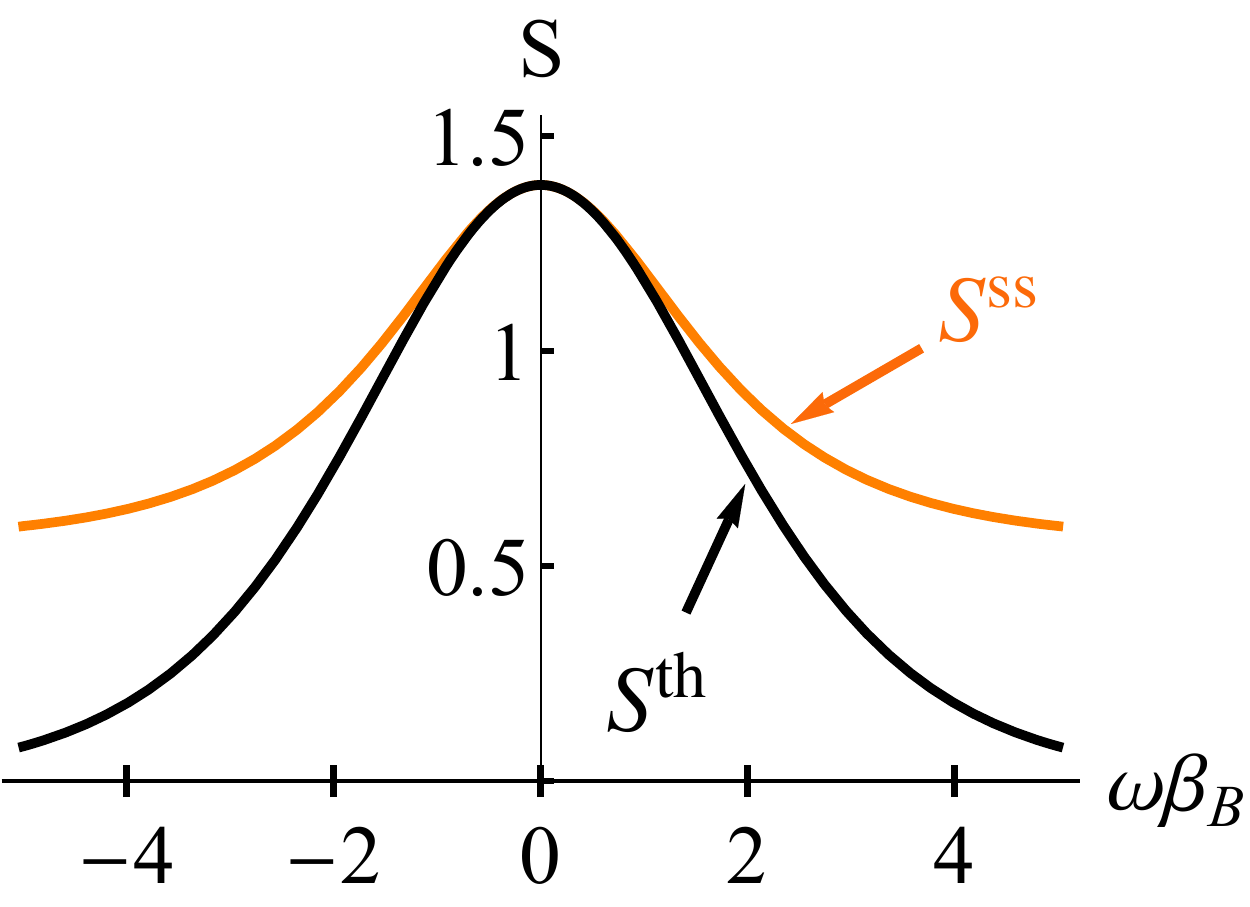}\includegraphics[width=4cm, height=3cm]{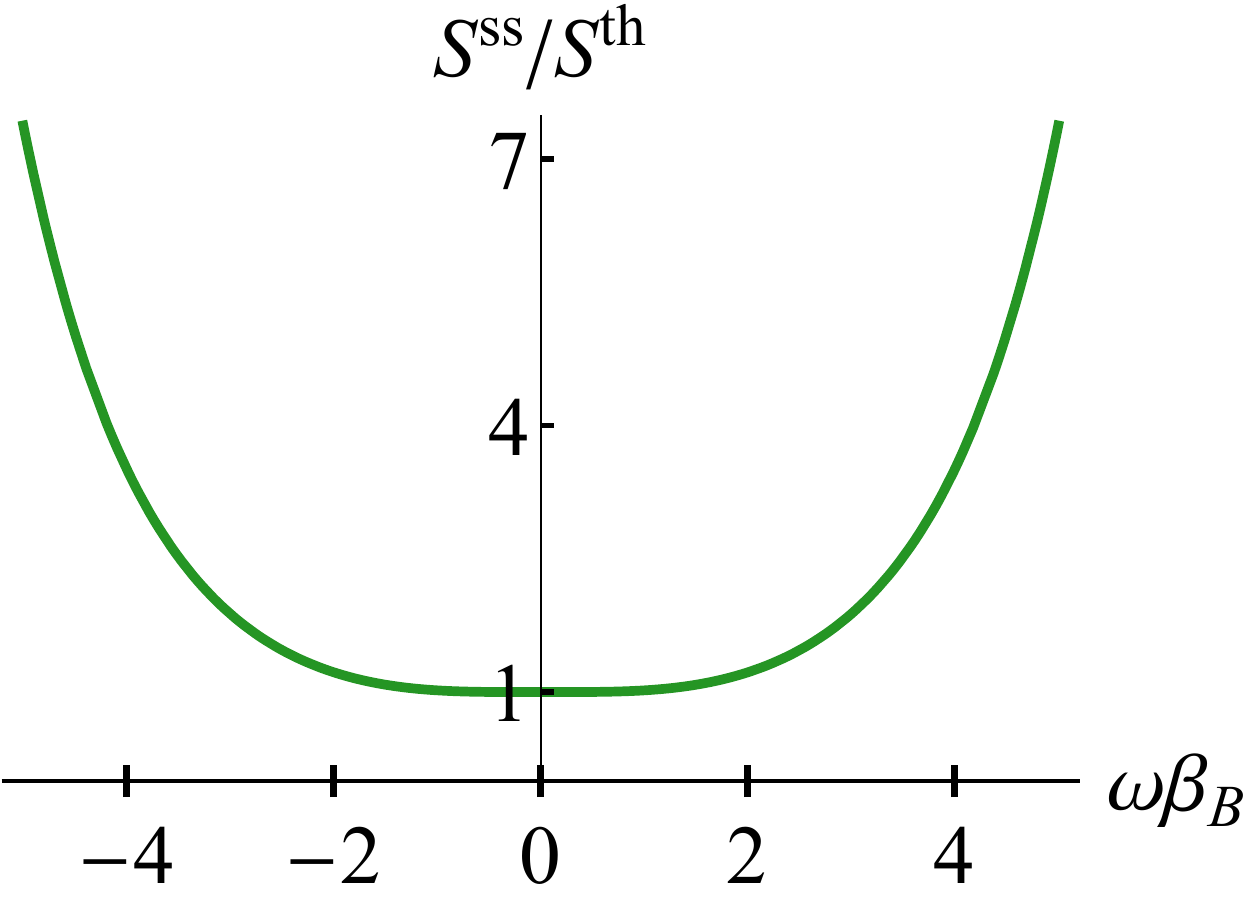}
(c)\includegraphics[width=4cm, height=3cm]{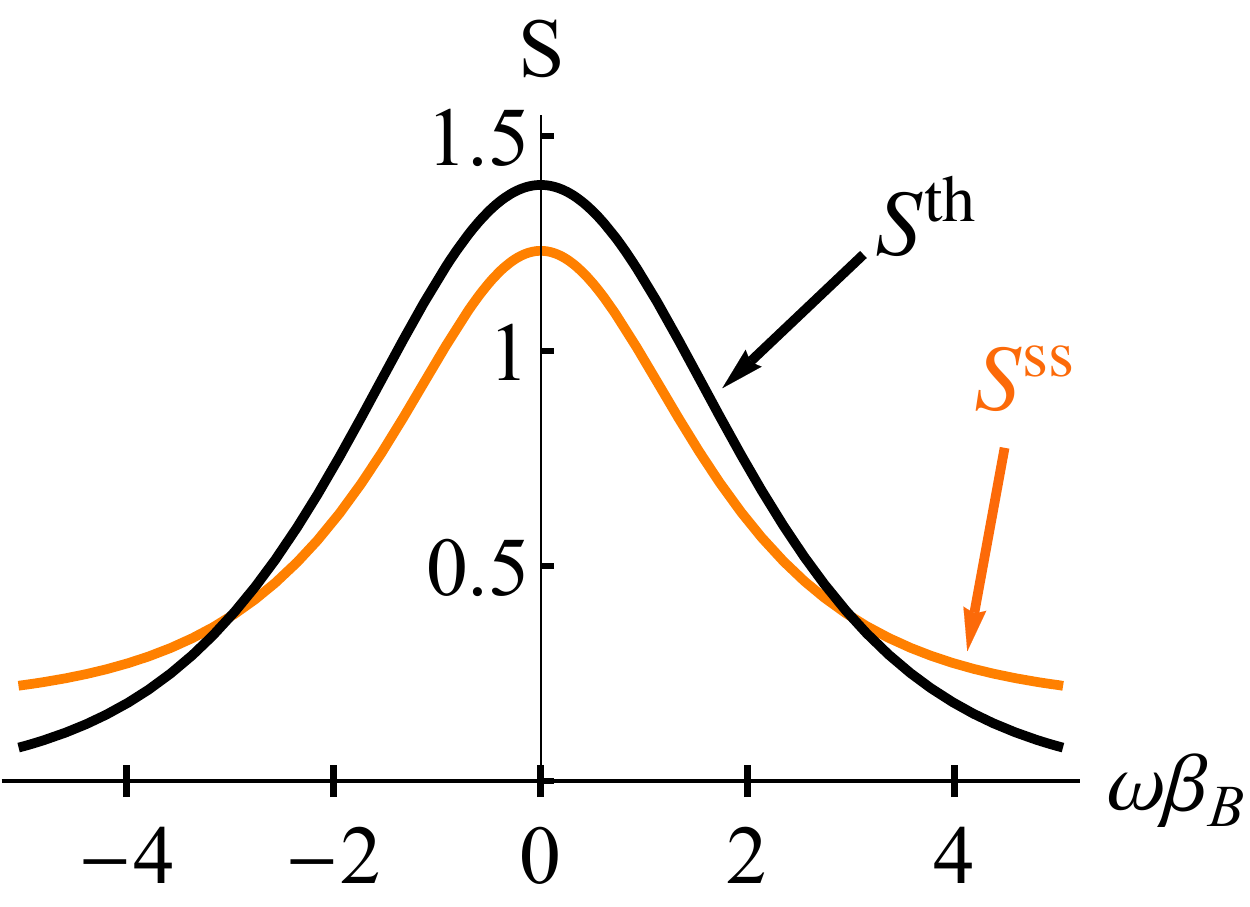}\includegraphics[width=4cm, height=3cm]{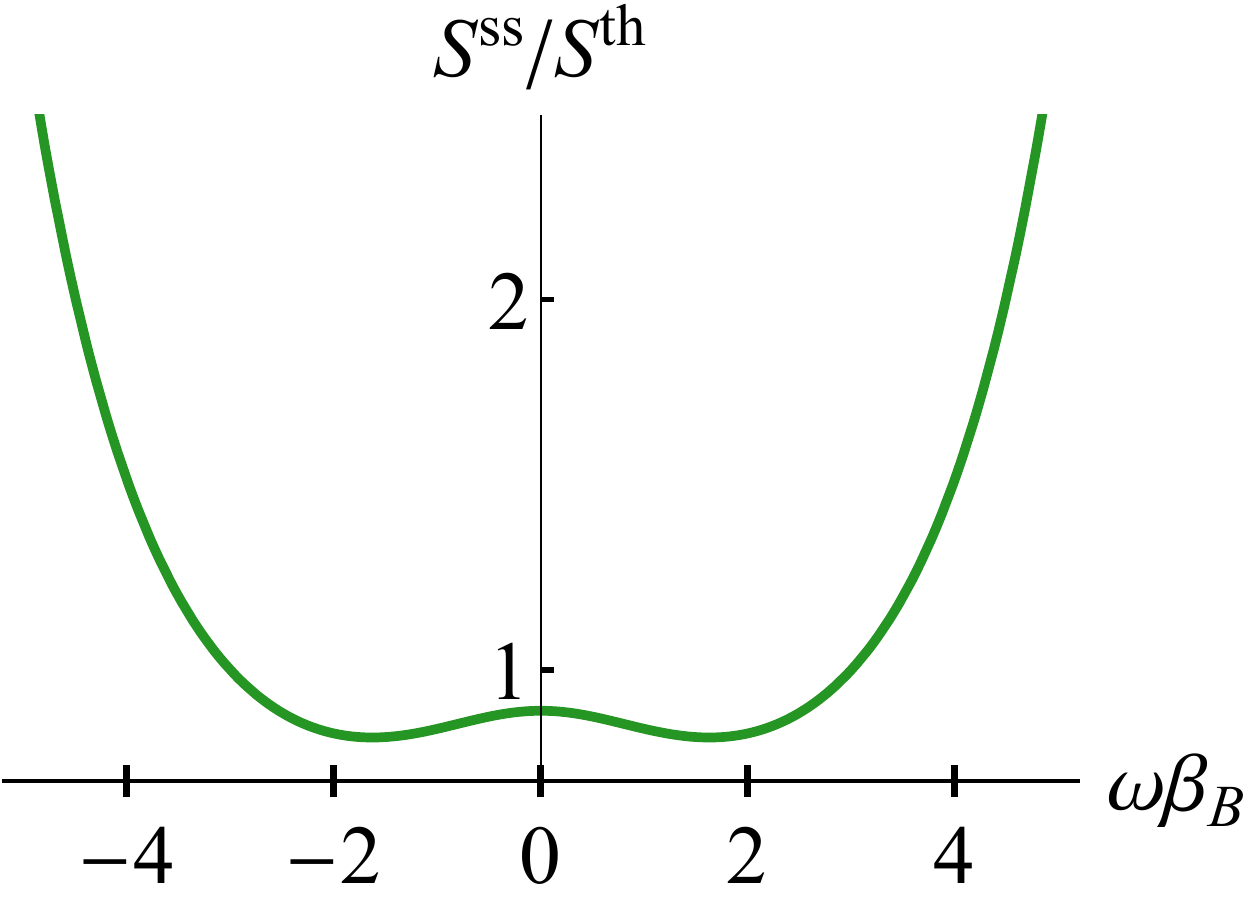}
\caption{Graphs of $S^{\rm ss}(\beta_B,\beta_0)$ (Orange curve), $S^{\rm th}(\beta_B)$ (Black curve), and $S^{\rm ss}(\beta_B,\beta_0)/S^{\rm th}(\beta_B)$ (Green curve, right panel) as a function of $\beta_B$ for (a) $\omega|\beta_0|\gg1$, (b) $\omega\beta_0=0$, and (c)  $\omega|\beta_0|=3$.}
\label{entropb0}
\end{figure}

  We now briefly show how strong the mitigation effect can be. The smallest values of $S^{\rm ss}(\beta_B,\beta_0)$ are obtained for $\omega|\beta_0| \gg 1$, reducing to the following expression,
\bea
S^{\rm ss}(\beta_B,\beta_0) &\rightarrow_{\omega|\beta_0| \gg 1}& S^{\rm th}(\beta_B) + \log r(\beta_B) \\
&&-\omega\beta_B \frac{1-e^{-\omega\beta_B}}{1+e^{-\omega\beta_B}} (r^{-1}(\beta_B)-1),\nn
\eea
which tends to $\frac{1}{2}S^{\rm th}(\beta_B)$ for $\omega |\beta_B| \gg1$ (see Fig \ref{entropb0} a). This is precisely the regime of low temperatures crucial in so many experiments and computational tasks. Then, for instance, a pair of two-level systems can be maintained in a state of energy and entropy up to twice smaller than the thermal energy $E^{\rm th}(\beta_B)$ and entropy $S^{\rm th}(\beta_B)$ only thanks to indistinguishability. 
This is also the regime of amplification of bath effects ($\beta_0<-\beta_B<0$ or $\beta_0>-\beta_B>0$) discussed in the previous Section \ref{sectionenergy}. Consequently, not only collective dissipation provides super cooling and super energy charging, but this is achieved with lower steady state entropy, which is always a highly desired in battery charging or refrigeration.

For sake of completeness we mention the regime $\omega|\beta_0| \ll 1$ where $S^{\rm ss}(\beta_B,\beta_0)$ takes its maximum values, given by the following expression,
\bea
&&S^{\rm ss}(\beta_B,\beta_0) \rightarrow_{\omega|\beta_0| \ll 1} S^{\rm th}(\beta_B) -\frac{3}{4}\log \frac{3}{4z(\beta_B)}\nn\\
&&\hspace{1.9cm} -\frac{1}{4} \log \left[ 1-e^{\omega\beta_B}Z_{+}(\beta_B)\left(\frac{3}{4z(\beta_B)} -1\right)\right] \nn\\
&&\hspace{1.9cm}-\omega\beta_B \frac{1-e^{-\omega\beta_B}}{1+e^{-\omega\beta_B}} \left(\frac{3}{4z(\beta_B)}-1\right),
\eea
which tends to $-\frac{1}{4}\log\frac{1}{4}-\frac{3}{4}\log\frac{3}{4}$  while $S^{\rm th}(\beta_B)$ tends to zero when $\omega|\beta_B| \gg1$. 
As illustrations, the graph of $S^{\rm ss}(\beta_B,\beta_0)$ as a function of $\omega\beta_0$  for $\omega \beta_B =2$ (or equivalently, for $\omega \beta_B=-2$) is shown in Fig. \ref{entropbb}, with the value of $S^{\rm th}(\beta_B)$ indicated for comparison. Graphs of $S^{\rm ss}(\beta_B,\beta_0)$, $S^{\rm th}(\beta_B)$, and $S^{\rm ss}(\beta_B,\beta_0)/S^{\rm th}(\beta_B)$ as functions of $\beta_B$ are shown in Fig. \ref{entropb0} for different value of $\omega\beta_0$. \\

\section{Conclusion}
Systems interacting with a bath can become effectively indistinguishable to that bath when they share the same characteristics with respect to the degrees of freedom that the bath is sensitive to. 
This was explicitly illustrated with the example of a pair of two-level atoms interacting with the free space electromagnetic field. 
One should also keep in mind that bath engineering can be helpful to reduce the bath sensitivity to some degrees of freedom and therefore increase the indistinguishability (as for instance adding a cavity field around a system to reduce the relevant electromagnetic modes \cite{Woods_2014}). 
The consequence of this indistinguishability is a collective dissipation (whose dynamics has been studied in superradiance problems). 

The study of the thermodynamic properties of the steady states generated by collective dissipation reveal the crucial role of the bath-induced coherences (fruit of indistinguishability). 
 They allow the system to reach an apparent temperature equal to the bath temperature while having lower (or higher) populations of excited levels than the thermal state $\rho^{\rm th}(\beta_B)$. In other words, without the bath-induced coherences the system could not have a steady state energy different from the thermal energy $E^{\rm th}(\beta_B)$. The result is either a mitigation or an amplification of the bath action, which manifests itself in both the energy and entropy of the system and can reach dramatic levels (up to $50\%$ of the thermal energy and entropy). In addition, the local steady states of the two-level systems are thermal states at temperatures $\beta_{\rm Loc}$ which can reach values much lower or higher than $\beta_B$ (up to $33\%$), providing one more dimension of the mitigation and amplification of the bath's action.
Such effects can be alternatively understood in terms of dark states. 

 The mitigation of the bath action can be useful for state preparation or protection for computational purposes or quantum error correction. 
 The amplification of the bath action suggests promising applications in charging of quantum batteries, thermal machines, and potentially also natural energy harvesting systems like photosynthesis. 
  In particular, the framework of autonomous thermal machines pioneered in \cite{Boukobza_2013,Gelbwaser_2014} and extended in \cite{autonomous} seems well fitted to apply the above results to charge quantum batteries, but also to cool them (loosing their role of proper battery to become systems to be refrigerated). This is particularly interesting since collective effects for cooling target systems have received little attention. 
 Furthermore, the phenomena described in this paper might be at the origin of the power increase witnessed in several papers on many-body quantum batteries \cite{Campaioli_2017,Ferraro_2018,Campaioli_2018} and on thermal machines with many-body working medium \cite{Wang_2009, Altintas_2014,Jaramillo_2016, Barrios_2017, Hardal_2017,Niedenzu_2018}, which would help pinpointing the resource at the origin of the power increase (still highly debated). This certainly deserves further studies. 
  We emphasise that the interesting and promising effects mentioned above emerge only thanks to the indistinguishability (of the two subsystems from the bath point of view). 
  
Throughout this paper we consider two idealised dynamics, corresponding either to totally distinguishable or totally indistinguishable subsystems. One can wonder for instance what happen when the two atoms are detuned. We expect, in the measure that the detuning is smaller than the inverse of the relaxation time (related to the strength of the bath coupling), that the above effects still hold but on a limited time interval given by the inverse of the detuning.     
Future studies should also investigate other relevant intermediary dynamics between totally distinguishable and totally indistinguishable subsystems.
 Still, numerous experiments already observed superradiance \cite{Skribanowitz_1973,Gross_1976,Rohlsberger_2010}, confirming that collective dissipation can be achieved in real system. Moreover, the present example with a pair of two-level systems reduces considerably the experimental challenges (in particular the interaction between subsystems does not break the indistinguishability) so that our results could be verified experimentally relatively easily.
 
  Finally, as already mentioned, this phenomenon is not limited to the present system. As long as the subsystems are indistinguishable to the bath, similar phenomena as the ones pointed out in this study should emerge. A future work will investigate the generalisation to larger systems, in particular ensembles of arbitrary number $n$ of arbitrary spin $s$, making the above results even more interesting and useful. \\
%
%

\acknowledgements
This  work  is  based  upon  research  supported  by  the
South  African  Research  Chair  Initiative  of  the  Department  of  Science  and  Technology  and  National  Research Foundation.

\appendix
\numberwithin{equation}{section}

\section{Distinguishable atoms}\label{apppol}
In this Section we evaluate the expectation value $ \langle B_i B_j \rangle_{\rho_{\rm bath}}$ in some specific situations, namely when the two atoms are far apart (having parallel polarisations) and when they have orthogonal polarisations while being at the same position. The free electromagnetic field at the point $\vec r$ decomposed in the plane wave basis can be written as \cite{Gross_1982} $\vec E(\vec r)= \vec E^{+}(\vec r) + \vec E^{-}(\vec r)$ with 
\be
\vec E^{+}(\vec r)=-i\sum_{\vec k,\vec \epsilon} \vec{\cal E}_{\vec k,\vec \epsilon} a_{\vec k,\vec \epsilon} e^{i\vec k.\vec r},
\ee
where $a_{\vec k, \vec \epsilon}$ is the annihilation operator associated to photons populating the planar mode of wave-vector $\vec k$ and polarisation $\vec \epsilon$. The vector $\vec {\cal E}_{\vec k, \vec \epsilon}=\sqrt{\frac{\hbar c k}{2\epsilon_0 V}} \vec \epsilon $ represents the electric field per photon ($V$ is an arbitrary volume of quantisation, much larger than the system). Therefore $\langle B_i B_j\rangle_{\rho_{\rm bath}}$ takes the following form,
\bea
\langle B_iB_j\rangle_{\rho_{\rm bath}} &=& d^2 \sum_{\vec k,\vec \epsilon} \vec e_i .\vec {\cal E}_{\vec k,\vec \epsilon}  \vec e_j.\vec {\cal E}_{\vec k,\vec \epsilon} e^{i\vec k.\vec r_{ij}} \langle a_{\vec k,\vec \epsilon} a^{\dag}_{\vec k,\vec \epsilon} \rangle_{\rho_{\rm bath}}\nn\\
&+& d^2 \sum_{\vec k,\vec \epsilon} \vec e_i .\vec {\cal E}_{\vec k,\vec \epsilon}  \vec e_j.\vec {\cal E}_{\vec k,\vec \epsilon} e^{-i\vec k.\vec r_{ij}} \langle a^{\dag}_{\vec k,\vec \epsilon} a_{\vec k,\vec \epsilon} \rangle_{\rho_{\rm bath}}\nn\\
\eea
where $\vec r_{ij} := \vec r_i -\vec r_j$, and assuming that the electromagnetic field is in a thermal state so that  $\langle a^{\dag}_{\vec k,\vec \epsilon} a_{\vec k',\vec \epsilon'} \rangle_{\rho_{\rm bath}} =\langle a_{\vec k,\vec \epsilon} a^{\dag}_{\vec k',\vec \epsilon'} \rangle_{\rho_{\rm bath}} = 0$ if $\vec k \ne \vec k'$ or $\vec \epsilon \ne \vec\epsilon'$. 
Transforming the discrete sum over the electromagnetic modes into a continuous one (volume integral) 
\be
\frac{1}{V} \sum_{\vec k,\vec \epsilon} \rightarrow \int \frac{d^3\vec k}{(2\pi)^3}\sum_{\vec\epsilon} = \frac{1}{(2\pi)^3}\int k^2dk \int d\Omega \sum_{\vec\epsilon},
\ee
where the integral over $\Omega$ denotes the integral over directions, one obtains
\bea
\langle B_iB_j\rangle_{\rho_{\rm bath}} &=& \frac{d^2}{(2\pi)^3}\int k^2dk \langle a_{k} a^{\dag}_{k} \rangle_{\rho_{\rm bath}}\int d\Omega e^{i\vec k.\vec r_{ij}} \nn\\
&&\hspace{1cm} \times \sum_{\vec \epsilon} \vec e_i .\vec {\cal E}_{\vec k,\vec \epsilon}  \vec e_j.\vec {\cal E}_{\vec k,\vec \epsilon}  \nn\\
&+& \frac{d^2}{(2\pi)^3}\int k^2dk \langle a^{\dag}_{k} a_{k} \rangle_{\rho_{\rm bath}}\int d\Omega e^{-i\vec k.\vec r_{ij}} \nn\\
&&\hspace{1cm} \times \sum_{\vec \epsilon} \vec e_i .\vec {\cal E}_{\vec k,\vec \epsilon}  \vec e_j.\vec {\cal E}_{\vec k,\vec \epsilon},  \nn\\
\eea
where we assumed that the occupation number $ \langle a^{\dag}_{\vec k,\vec \epsilon} a_{\vec k,\vec \epsilon} \rangle_{\rho_{\rm bath}}$ of each electromagnetic modes does not depend on the polarisation $\vec \epsilon$ neither on the direction of $\vec k$ but only on the norm of the wave-vector $k$. \\

{\bf Parallel polarisation}. When $\vec e_i =\vec e_j$, 
\bea
\int d\Omega e^{i\vec k.\vec r_{ij}}  \sum_{\vec \epsilon} \vec e_i .\vec {\cal E}_{\vec k,\vec \epsilon}  \vec e_j.\vec {\cal E}_{\vec k,\vec \epsilon} &=& \int d\Omega e^{i\vec k.\vec r_{ij}}  \sum_{\vec \epsilon} (\vec e_i .\vec {\cal E}_{\vec k,\vec \epsilon} )^2 \nn\\
&=& F_{ij}(kr_{ij})
\eea
where $F_{ij}(kr_{ij})$ is the function defined in \cite{Gross_1982} which depends on $k, r_{ij}$ and $\vec e_i$. The detail of the expression of $F_{ij}(kr_{ij})$ is not important in our present analysis. However, what is important is that $F_{ij}(kr_{ij})$ tends to zero when $kr_{ij}$ is much larger than 1.
Taking into account that only the resonant electromagnetic modes ($ k c = \omega$) interact predominantly with the atoms, we can consider that $\langle B_iB_j\rangle_{\rho_{\rm bath}} = 0$ when $r_{ij}$ is much larger than $\lambda_a = c/\omega$, which corresponds to the emission wavelength, as announced in the main text. \\

{\bf Orthogonal polarisations}. We now consider that the two atoms are confined in a volume much smaller than the emission wavelength $c/\omega$. Therefore, for the modes interacting predominantly with the atoms we can consider that $kr_{ij} \simeq 0$ so that the expectation value is reduced to
\bea
\langle B_iB_j\rangle_{\rho_{\rm bath}} &=& \frac{d^2}{(2\pi)^3}\int k^2dk \langle a_{k} a^{\dag}_{k} \rangle_{\rho_{\rm bath}} \nn\\
&&\hspace{1cm} \times \int d\Omega \sum_{\vec \epsilon} \vec e_i .\vec {\cal E}_{\vec k,\vec \epsilon}  \vec e_j.\vec {\cal E}_{\vec k,\vec \epsilon}  \nn\\
&+& \frac{d^2}{(2\pi)^3}\int k^2dk \langle a^{\dag}_{k} a_{k} \rangle_{\rho_{\rm bath}} \nn\\
&&\hspace{1cm} \times \int d\Omega\sum_{\vec \epsilon} \vec e_i .\vec {\cal E}_{\vec k,\vec \epsilon}  \vec e_j.\vec {\cal E}_{\vec k,\vec \epsilon} ~.  \nn\\
\eea
Remembering that for each wave-vector $\vec k$ the sum of $\vec \epsilon$ runs over any two orthonormal vectors belonging to the plan perpendicular to $\vec k$, we have that $\vec e_{i}.\vec e_{j} = e_{i,\vec k}e_{j,\vec k} +\sum_{\vec \epsilon} \vec e_i .\vec {\cal E}_{\vec k,\vec \epsilon}  \vec e_j.\vec {\cal E}_{\vec k,\vec \epsilon}$, where $e_{i,\vec k} : = \vec e_i.\vec k/k$. Assuming that the polarisations of the two atoms are orthogonal we have $\vec e_{i}.\vec e_{j}=0$ so that $\sum_{\vec \epsilon} \vec e_i .\vec {\cal E}_{\vec k,\vec \epsilon}  \vec e_j.\vec {\cal E}_{\vec k,\vec \epsilon}=-e_{i,\vec k}e_{j,\vec k} $. 
  Therefore, it follows 
\bea
\int d\Omega\sum_{\vec \epsilon} \vec e_i .\vec {\cal E}_{\vec k,\vec \epsilon}  \vec e_j.\vec {\cal E}_{\vec k,\vec \epsilon} &=& \int d\Omega  e_{i,\vec k} e_{j,\vec k}\nn\\
&=& \int_{S} dS xy  
\eea
where $S$ is the unit sphere (sphere centered at $k=0$ with radius 1), and $x$ and $y$ stand for the Cartesian coordinates (defined along the axis $\vec e_i$ and $\vec e_j$, so that $x:=e_{i,\vec k}$, and $y:=e_{j,\vec k}$) of a point on the sphere $S$. Such integral is equal to zero implying that the expectation value $\langle B_iB_j\rangle_{\rho_{\rm bath}} $ is null.

\section{Interaction between the two confined atoms}\label{tlatoms}
The expression of the coupling constant related to the interaction between the two confined atoms is given by \cite{Gross_1982} $\Omega_{I}=\frac{d^2}{4\pi\epsilon_0r_{12}^3}\left[1-\frac{3(\vec{\epsilon}_i.\vec r_{12})^2}{r_{12}^2}\right]$. The interaction corresponds to the Van der Waals interaction between the two atoms at position $\vec r_1$ and $\vec r_2$, with $\vec r_{12}:=\vec r_1 -\vec r_2$, and $\vec \epsilon_i$ is the polarisation of the electric dipole (assumed to be the same for both atoms). In particular, if the atoms are far apart their interaction vanishes.

\section{Steady state of a pair of indistinguishable two-level systems}\label{dynamics}
In this Section we derive the steady state of the collective dissipation described by a generalisation of the master equation Eq. \eqref{meind} of the main text,
\bea\label{indsm}
\dot{\rho}_S^I &=& -i\Omega_L \sum_{i=1}^2 [\sigma_i^{+}\sigma_i^{-},\rho_S^I]-i\Omega_{I}[\sigma_1^{+}\sigma_{2}^{-}+\sigma_1^{-}\sigma_{2}^{+},\rho_S^I] \nn\\
&&+ G(\omega) (2S^-\rho_S^I S^+ -S^+S^-\rho_S^I - \rho_S^IS^+S^-)\nn\\
&&+ G(-\omega) (2S^+\rho_S^I S^- -S^-S^+\rho_S^I - \rho_S^IS^-S^+),\nn\\
\eea
where the dissipation rate $G(\omega)$ and the pumping rate $G(-\omega)$ depend on the characteristics of the bath or effective bath (which can be the result of the interaction of several baths or collisional model). This amounts to replace $g[n(\omega)+1]$ by $G(\omega)$ and $gn(\omega)$ by $G(-\omega)$. One should note that the temperature (or apparent temperature \cite{apptemppaper}) of the effective bath is given by $e^{\omega \beta_B} = G(\omega)/G(-\omega)$ which can be larger or smaller than 1, corresponding to $\beta_B$ positive or negative. 
 The dynamics can be easily solved by considering the basis $\{|\psi_0\ket,|\psi_+\ket,|\psi_-\ket,|\psi_1\ket\}$ with $|\psi_{\pm}\ket =(|0\ket|1\ket\pm|1\ket|0\ket)/\sqrt{2}$, $|\psi_{0}\ket = |0\ket|0\ket$, and $|\psi_1\ket=|1\ket|1\ket$. In such basis the collective ladder operators can be expressed as $S^{+}=\sqrt{2}|\psi_+\ket \bra\psi_0|+\sqrt{2}|\psi_1\ket\bra\psi_+|$ and $S^{-}=\sqrt{2}|\psi_0\ket \bra\psi_{+}|+\sqrt{2}|\psi_{+}\ket\bra\psi_1|$. From \eqref{indsm} one obtains the following dynamics for the populations $p_i:=\bra \psi_i|\rho_S|\psi_i\ket$, $i=0,1,+,-$, 
 \bea\label{sys}
 &&\dot{p}_1 = 4 G(-\omega)p_+ - 4G(\omega)p_1\nn\\
 &&\dot{p}_0=4G(\omega)p_+ -4G(-\omega)p_0\nn\\
 &&\dot{p}_+ = 4G(\omega)(p_1-p_+)+4G(-\omega)(p_0-p_+)\nn\\
 &&\dot{p}_- = 0.
 \eea 
 
 The steady state populations can be obtained by canceling the time derivatives in the above system of equations. Alternatively, one can also solve the above system. This is simplified by noting that $\dot{p}_1+\dot{p}_0+\dot{p}_+ = 0$, which implies that $r:=p_1+p_0+p_+$ is a constant determined by the initial conditions. The system can therefore be reduced to a system of two linearly independent equations (substituting for instance $p_1$ by $r-p_0-p_+$),
\bea
 &&\dot{p}_0=4G(\omega)p_+ -4G(-\omega)p_0\nn\\
 &&\dot{p}_+ = -4[G(\omega)-G(-\omega)]p_0-4[2G(\omega)+G(-\omega)]p_+ \nn\\
 &&\hspace{0.9cm}+ 4G(\omega)r.
 \eea 
The reduced system is diagonalised by the quantities $q^{\pm} := p_+ + (1\pm\sqrt{G(-\omega)/G(\omega)})p_0$, with the associated eigenvalues $a^{\pm}:= 4[ \pm \sqrt{G(\omega)G(-\omega)}-G(\omega)-G(-\omega)]$, so that 
\be 
\dot q^{\pm} = a^{\pm} q^{\pm} + 4G(\omega)r,
\ee
and 
\be
q^{\pm}(t)= e^{a^{\pm} t} q^{\pm}(0) +4G(\omega)r\frac{e^{a^{\pm}t}-1}{a^{\pm}}.
\ee
From the time evolution of $q^{\pm}(t)$ one obtains straightforwardly the expression for the time evolution of the populations $p_0$, $p_+$, and $p_1$.   
Using any of the above methods, the steady state populations are found to be 
\bea
&&p_0^{\rm ss}=rZ_{+}^ {-1}(\beta_B),\nn\\
&& p_+^{\rm ss}=rZ_{+}^{-1}(\beta_B)e^{-\omega \beta_B},\nn\\ 
&& p_1^{\rm ss}= rZ_{+}^{-1}(\beta_B)e^{-2\omega \beta_B},\nn\\
&& p_{-}^{\rm ss} = p_{-}(t=0)=1-r,
 \eea
 with $Z_{+}(\beta_B) := 1 +e^{-\omega\beta_B}+e^{-2\omega\beta_B}$. \\

For the coherences, defined as $\rho_{ij}:=\bra \psi_i|\rho_S^I|\psi_j\ket$, $i,j \in \{0,1,+,-\}$, one obtains (including the Lamb shift in the interaction picture),
\bea 
&&\dot{\rho}_{+,-} = -2\big[G(\omega)+G(-\omega) + i\Omega_{I} \big] \rho_{+,-}\nn\\
&&\dot{\rho}_{1,-} = -\big[2G(\omega)+i\Omega_{I}\big]\rho_{1,-} \nn\\
&&\dot{\rho}_{0,-} = -\big[2G(-\omega)+i\Omega_{I}\big]\rho_{0,-} \nn\\
&&\dot{\rho}_{1,0} = -2[G(\omega)+G(-\omega)]\rho_{1,0}
\eea
which straightforwardly gives $0$ as steady state solution. The dynamics of the two remaining coherences is coupled,
\bea
&&\dot{\rho}_{1,+} = -\big[2(2G(\omega)+G(-\omega))-i\Omega_{I}\big]\rho_{1,+} + 4G(-\omega)\rho_{+,0}\nn\\
&&\dot{\rho}_{+,0} =  -\big[2g(G(\omega)+2G(-\omega))+i\Omega_{I}\big]\rho_{+,0} + 4G(\omega)\rho_{1,+},\nn\\
\eea
and also leads to $0$ as steady state solution. Finally, one can write the steady state in the form,
\bea
 && \rho^{\rm ss}(\beta_B,r) := (1-r)|\psi_{-}\ket\bra\psi_{-}| + rZ_{+}^{-1}(\beta_B) \nn\\
&&\hspace{0.5cm}\times \Big(e^{-2\omega\beta_B}|\psi_1\ket\bra \psi_1|+e^{-\omega\beta_B}|\psi_+\ket\bra\psi_+| + |\psi_0\ket\bra\psi_0|\Big),\nn\\
\eea
as announced in the main text.\\

\section{The role of the indistinguishability from the bath}\label{appindistin}
In this Section we come back on the role of the indistinguishability from the bath. The underlying mechanism can be generalise in the following way. 
Let's consider a system initially in a pure state $|a\ket$. This system undergoes a process (enters a black box) with two different outputs $|b\ket$ and $|c\ket$ (orthogonal states) of same probability $1/2$. We assume that the system's bath, or more generally the surrounding of the system, does not distinguish whether the system is in the state $|b\ket$ or $|c\ket$ (we give some example of such situations in the following). This can be alternatively formulated in the following way: no information about the actual system's state is leaked to the bath or surrounding environment. If such conditions are fulfilled, the system is left after the process (black box) in a coherent superposition of $|b\ket$ and $|c\ket$. 

This phenomenon is well-known in quantum optics for the design of interferometers and in the double slit experiment. 
 In both experiments, the incident photon goes through a process (the double slit or the beam splitter). The two outcomes are two different paths/modes. If no information about the path used by the photon is leaked to the surrounding environment, the photon is in a coherent superposition of paths which interfere. Conversely, the more information is available about the path used by the photon the less coherent is the superposition and the weaker are the interferences. Thus, the indistinguishability of the two paths from the point of view of the environment enables the coherent superposition of the two paths.   
 
The absorption of one bath excitation by the pair of atoms follows the same scenario. To draw a simple comparison, let's assume both atoms are initially in the ground state. Then, they go through the process which is the absorption of one bath excitation. There are two orthogonal outputs, $|b\ket \equiv$ {\it atom 1 in the excited state and atom 2 in the ground state} and $|c\ket \equiv$ {\it atom 1 in the ground state and atom 2 in the excited state}. The indistinguishability of the two atoms from the bath's point of view means that the bath cannot distinguish which atom absorbs the excitation. Therefore, the output states $|b\ket$ and $|c\ket$ are indistinguishable for the bath, so that the process generates a coherent superposition of $|b\ket$ and $|c\ket$. 

Of course this toy model can be extended to situations where there are more than two output states, where the output probabilities are unbalanced, and where partial information is leaked to the bath (reducing the coherence of the superposition).
Importantly, we stress that the object/system performing the process does not need to contain any quantum features. The double slit and the beam splitter are classical objects. 
Similar ideas based on indistinguishability have been exploited in \cite{Lofranco_2016,Lofranco_2018,Kysela_2019} for entanglement generation.

\section{Local Temperature}\label{localtemp}
\begin{figure}
\centering
(a)\includegraphics[width=6cm, height=4cm]{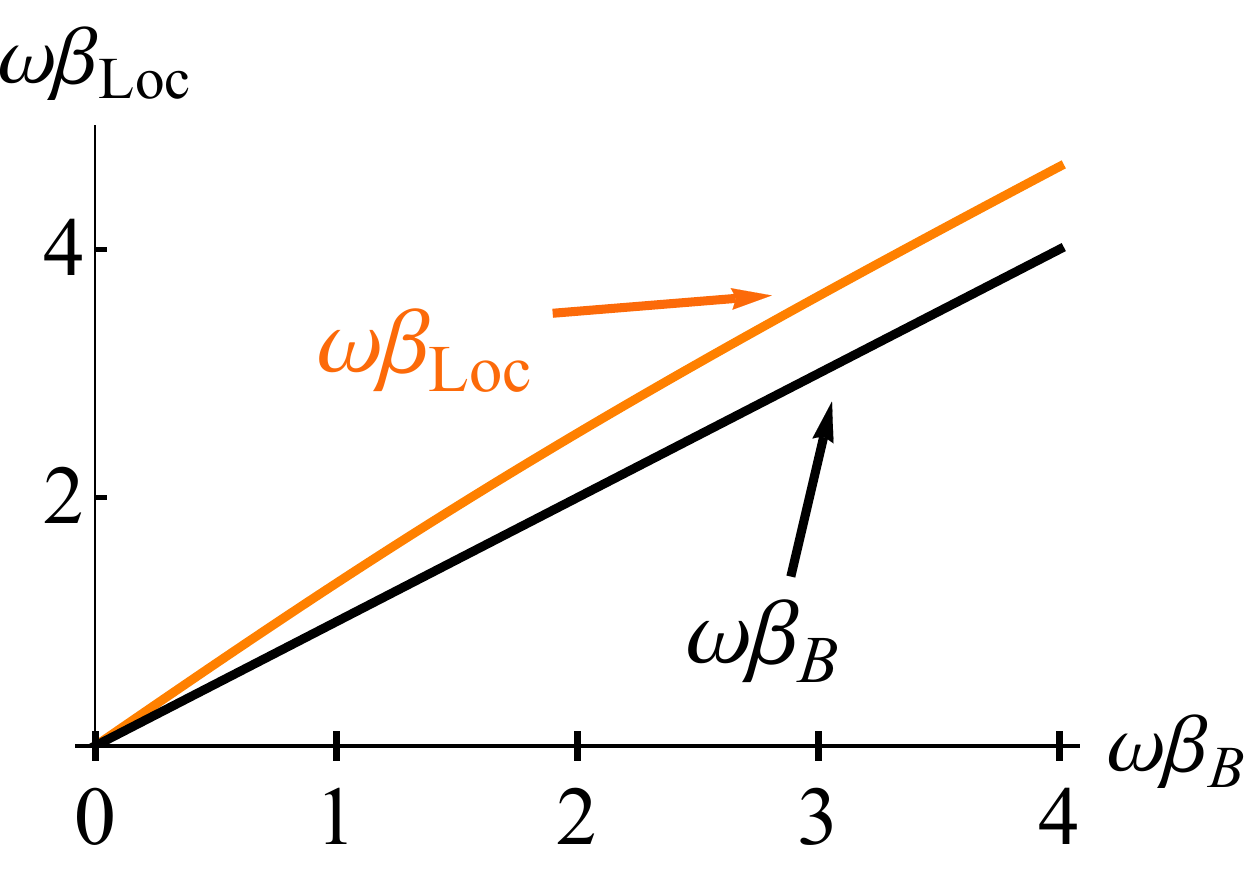}
(b)\includegraphics[width=6.7cm, height=4.3cm]{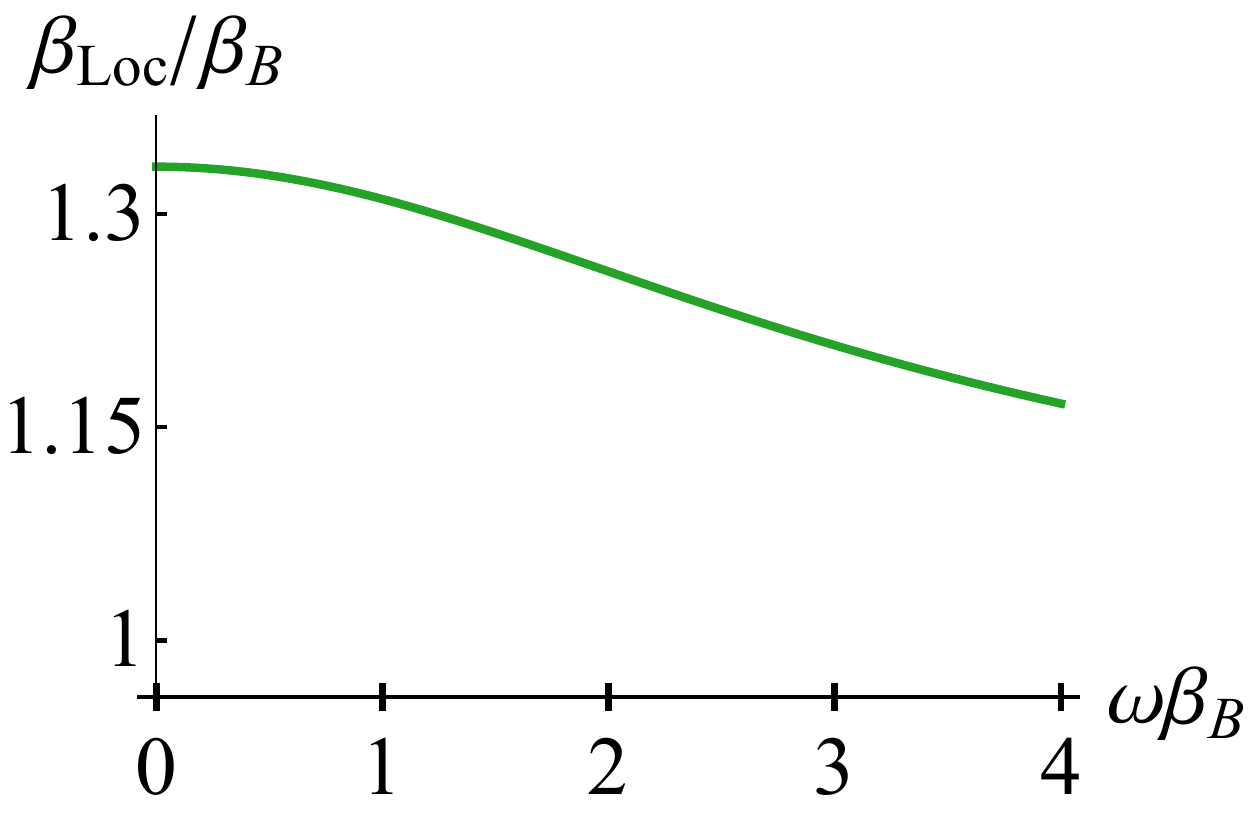}
\caption{(a) Graph of the steady state local inverse temperature $\beta_{\rm Loc}$ (Orange curve) and as a function of the bath inverse temperature $\omega \beta_B \in [0;4]$  for $\omega|\beta_0| \gg1$. As a comparison the bath inverse temperature is indicated by the Black curve. (b) Graph of the ratio $\beta_{\rm Loc}/\beta_B$ as a function of $\omega \beta_B$.}
\label{loctemp}
\end{figure}

In this Section we provide graphs of the steady state local temperature in function of the bath temperature. The first graph Fig. \ref{loctemp} shows how much the cooling can be amplified when the pair of two-level systems is initialised with highly inverted population, reminiscent of the Mpemba effect \cite{Mpemba_1969,Lasanta_2017,Lu_2017} (expect that here hotter initial states reach colder temperatures). Similar graphs can be obtained showing the mitigation effects. 

\section{Apparent temperature of the steady states}\label{smapptempss}
Inserting the expression of the steady state \eqref{genss} into the definition of the apparent temperature \eqref{defapptemp} one obtains
\bea
{\cal T} &=& \omega \left(\log\frac{\bra \psi_0|\rho^{\rm ss}(\beta_B,r)|\psi_0\ket +\bra \psi_+|\rho^{\rm ss}(\beta_B,r)|\psi_+\ket}{\bra \psi_1|\rho^{\rm ss}(\beta_B,r)|\psi_1\ket +\bra \psi_+|\rho^{\rm ss}(\beta_B,r)|\psi_+\ket} \right)^{-1}\nn\\
\eea
with $\bra\psi_0|\rho^{\rm ss}(\beta_B,r)|\psi_0\ket = rZ_{+}^{-1}(\beta_B)$, $\bra\psi_+|\rho^{\rm ss}(\beta_B,r)|\psi_+\ket = rZ_{+}^{-1}(\beta_B)e^{-\omega\beta_B}$, and $\bra\psi_1|\rho^{\rm ss}(\beta_B,r)|\psi_1\ket = rZ_{+}^{-1}(\beta_B)e^{-2\omega\beta_B}$. As a result,  
\be
{\cal T} = 1/\beta_B.
\ee

\section{Behaviour of the steady state entropy in term of the coherence}\label{appentropy}
\begin{figure}
\centering
\includegraphics[width=7cm, height=4.5cm]{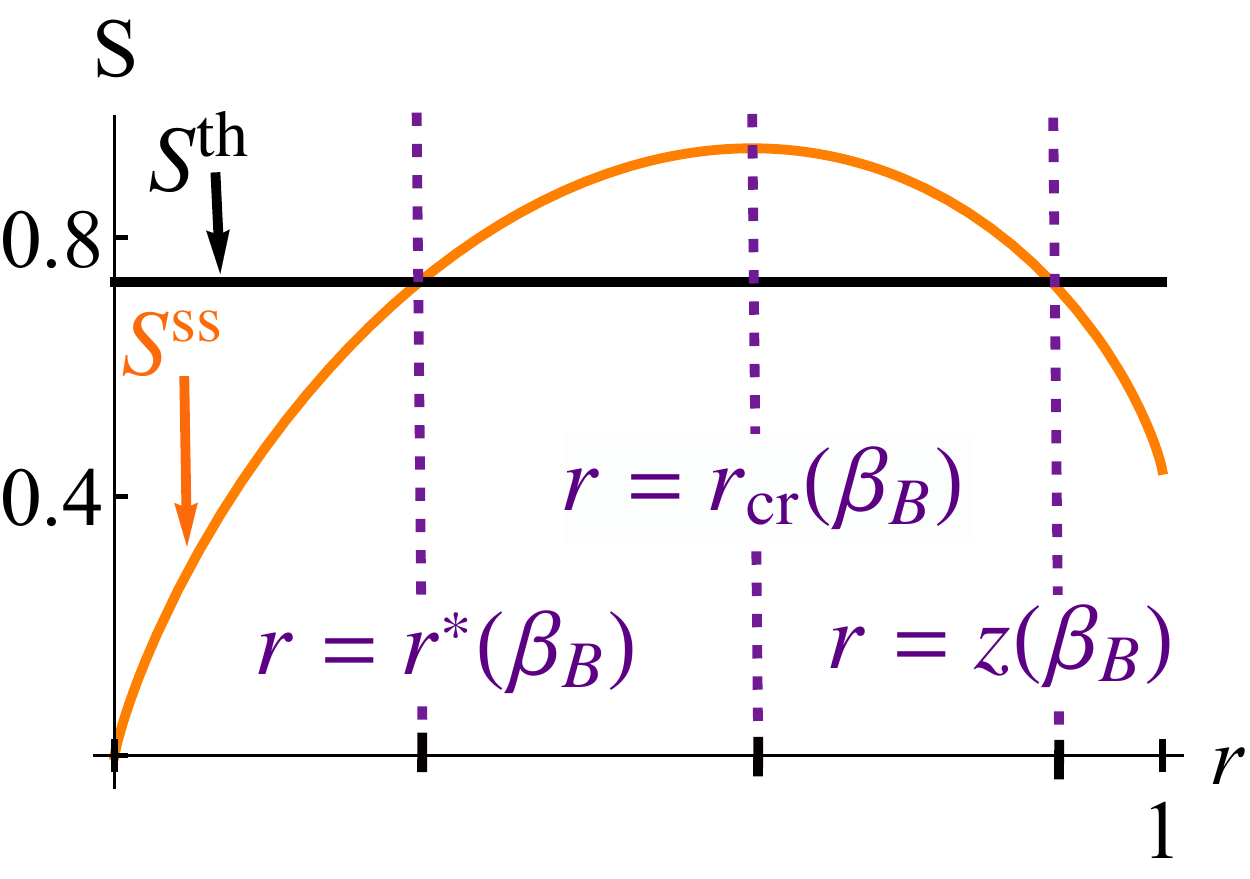}
\caption{Graphs of $S^{\rm ss}(\beta_B,r)$ (Orange curve) as a function of $r$ for $\beta_B=2$. The value of $S^{\rm th}(\beta_B)$ is indicated by the Black line. Indicated by the dashed lines, the maximum of $S^{\rm ss}(\beta_B,r)$ at the point $r=r_{\rm cr}(\beta_B)$, and the two points such that $S^{\rm ss}(\beta_B,r)=S^{\rm th}(\beta_B)$ at $r=z(\beta_B)$ and $r=r^{*}(\beta_B)$. }
\label{entropr}
\end{figure}

In this Section we study the behaviour of the steady state entropy in term of the coherence. Note that for a fixed bath temperature, the steady state coherence is entirely determined by $r$. Using Eq. \eqref{entropyc} we compute the derivative of the steady state entropy with respect to $r$ and obtain, 
\bea
\frac{\partial S^{\rm ss}(\beta_B,r)}{\partial r}&=& \log \left(\frac{1}{r} -1\right) + \log Z_{+}(\beta_B)\nn\\ &&+\omega\beta_B\frac{(e^{-\omega\beta_B}+2e^{-2\omega\beta_B})}{Z_{+}(\beta_B)}.
\eea
The derivative $\frac{\partial S^{\rm ss}(\beta_B,r)}{\partial r}$ is positive for $r$ in $[0;r_{cr}(\beta_B)]$ and negative on $[r_{cr}(\beta_B),1]$, with 
\be
r_{cr}(\beta_B) := \frac{z(\beta_B)}{1 + \frac{e^{-\omega^{*}\beta_B}-e^{-\omega\beta_B}}{Z(\beta_B)}},
\ee
and $\omega^{*}:=\omega \frac{e^{-\omega\beta_B}+2e^{-2\omega\beta_B}}{Z_{+}(\beta_B)}$. The graph of $S^{\rm ss}(\beta_B,r)$ as a function of $r$ is shown in Fig. \ref{entropr} (for $\omega\beta_B=2$). Note that $r_{cr}(\beta_B) \leq z(\beta_B) $ for any $\beta_B\ne 0$ (equality only for $\beta_B=0$). Since $S^{\rm ss}(\beta_B,\beta_B) = S^{\rm th}(\beta_B)$ when $r=z(\beta_B)$ (equivalent to $c=0$), we have that $S^{\rm ss}(\beta_B,r) <S^{\rm th}(\beta_B)$ for $c>0$. Furthermore, given that $\frac{\partial S^{\rm ss}(\beta_B,r)}{\partial r}$ is positive for $r\leq r_{cr}(\beta_B)$, and $S^{\rm ss}(\beta_B,r=0)=0$ while $S^{\rm ss}(\beta_B,r)\geq S^{\rm th}(\beta_B)$ for $r=r_{cr}(\beta_B)$, there is a value $r$ belonging to the interval $]0;r_{cr}(\beta_B)]$ such that $S^{\rm ss}(\beta_B,r)=S^{\rm th}(\beta_B)$. We denote by $r^{*}(\beta_B)$ such value, and by $c^{*}(\beta_B)$ the corresponding value of the coherence ($c^{*}(\beta_B) := r^{*}(\beta_B)/r(\beta_B) -1 \leq r_{cr}(\beta_B)/r(\beta_B) -1)$. Therefore, $S^{\rm ss}(\beta_B,r)>S^{\rm th}(\beta_B)$ for $c \in ]c^{*}(\beta_B);0[$, and $S^{\rm ss}(\beta_B,r)<S^{\rm th}(\beta_B)$ for $c \in [-1;c^{*}(\beta_B)[$ (as announced in the main text).

Considering only initial thermal states, the parameter $r$ becomes equal to $z(\beta_B)$ which takes value in the interval $[\frac{3}{4};1]$. On the other hand, one can verify that $r_{\rm cr}(\beta_B)\leq 3/4$ for any $\beta_B$. This guarantees that $S^{\rm ss}(\beta_B,\beta_0)$ stays larger than $S^{\rm th}(\beta_B)$ for any initial inverse temperature $\beta_0$ such that $|\beta_0|<|\beta_B|$ (equivalent to $z(\beta_0)<z(\beta_B)$).
We finally reach the statement made in the main text: $S^{\rm ss}(\beta_B,\beta_0)$ is strictly larger (smaller) than $S^{\rm th}(\beta_B)$ if and only if $|\beta_0| <|\beta_B|$ ($|\beta_0|>|\beta_B|$).

\end{document}